\documentclass[12pt,letterpaper]{article}
\pdfoutput=1

\usepackage{amsmath,amssymb,calc, amsthm,bbm, epsfig,psfrag, mathtools}
\usepackage{graphicx, enumerate}
\usepackage[numbers,sort&compress]{natbib}
\usepackage[dvipsnames]{xcolor}
\usepackage{tensor}

\usepackage[utf8]{inputenc} 
\allowdisplaybreaks

\newcommand{\Comment}[1]{{}}
\definecolor{darkblue}{rgb}{0.15,0.35,0.55}
\definecolor{reddish}{rgb}{0.65, 0.2, 0.2}
\usepackage[linktocpage=true]{hyperref}
\hypersetup{
colorlinks=true,
citecolor=darkblue,
linkcolor=reddish,
urlcolor=darkblue,
pdfauthor={},
pdftitle={},
pdfsubject={}
}

\makeatletter
\renewcommand\section{\@startsection {section}{1}{\z@}%
                                   {-3.5ex \@plus -1ex \@minus -.2ex}%nn
                                   {2.3ex \@plus.2ex}%
                                   {\normalfont\large\bfseries}}
\renewcommand\subsection{\@startsection{subsection}{2}{\z@}%
                                     {-3.25ex\@plus -1ex \@minus -.2ex}%
                                     {1.5ex \@plus .2ex}%
                                     {\normalfont\bfseries}}
\makeatother

\let\non\nonumber

\def\bea#1\eea{\begin{align}#1\end{align}}

\def\bes #1\ees{\begin{split}#1\end{split}}

\newcommand{\be}{\begin{equation}}
\newcommand{\ee}{\end{equation}}

\newcommand{\bma}{\begin{pmatrix}}
\newcommand{\ema}{\end{pmatrix}}

\newcommand{\C}[1]{$(\ref{#1})$}

\newfont{\goth}{ygoth.tfm scaled 1200}                   % gothic font (usual)
 \numberwithin{equation}{section}

\def\1{{(1)}}
\def\2{{(2)}}
\def\3{{(3)}}

\def\s{\sigma}

\newcommand\Tb{\overline{T}}

\newcommand{\ppp}{{+++}} 
\newcommand{\mmm}{{---}}
\newcommand{\mmp}{{--+}}
\newcommand{\ppm}{{++-}}
\newcommand{\pppp}{{++++}}
\newcommand{\mmmm}{{----}}
\newcommand{\ppmm}{{++--}}
\newcommand{\mmpp}{{--++}}
\newcommand{\pp}{{++}}
\newcommand{\mm}{{--}}
\newcommand{\tp}{\theta^+}
\newcommand{\tm}{\theta^-}
\newcommand{\ww}{{\pm\pm}}
\newcommand{\www}{{\pm\pm\pm}}
\newcommand{\wwww}{{\pm\pm\pm\pm}}

\newcommand{\xp}{x^\prime}
\newcommand{\lrb}{\left(}
\newcommand{\rrb}{\right)}
\newcommand{\ld}{\left\langle}
\newcommand{\rd}{\right\rangle}
\newcommand{\T}{\mathcal{T}}

\begin{document}
\begin{titlepage}

\begin{center}

%\today 
{November 5, 2018}
\hfill         \phantom{xxx}  EFI-18-17

\vskip 2 cm {\Large \bf Supersymmetry and $T \Tb$ Deformations} 
\vskip 1.25 cm {\bf Chih-Kai Chang, Christian Ferko and Savdeep Sethi}\non\\
\vskip 0.2 cm
 {\it Enrico Fermi Institute \& Kadanoff Center for Theoretical Physics \\ University of Chicago, Chicago, IL 60637, USA}

\vskip 0.2 cm
%{ Email:} \href{mailto:ckchang@uchicago.edu}{ckchang@uchicago.edu}, \href{mailto:cferko@uchicago.edu}{cferko@uchicago.edu}, \href{mailto:sethi@uchicago.edu}{sethi@uchicago.edu}

\end{center}
\vskip 1.5 cm

\begin{abstract}
\baselineskip=18pt
    
We propose a manifestly supersymmetric generalization of the solvable $T \Tb$ deformation of two-dimensional field theories. For theories with $(1,1)$ and $(0,1)$ supersymmetry, the deformation is defined by adding a term to the superspace Lagrangian built from %bilinears of 
a superfield containing the supercurrent. 
%??? This deformation is defined for $(1,1)$ and $(0,1)$ theories by adding a term in the superspace Lagrangian which ??? built from bilinears of a superfield containing the supercurrent ??? can be interpreted as the square of the supercurrent superfield. 
We prove that the energy levels of the resulting deformed theory are determined exactly in terms of those of the undeformed theory. This supersymmetric deformation extends to higher dimensions, where we conjecture that it might provide a higher-dimensional analogue of $T \Tb$, producing supersymmetric Dirac or Dirac-Born-Infeld actions in special cases. 

%. For special cases, this deformation gives the supersymmetric Dirac or Dirac-Born-Infeld actions. 

%, and we conjecture that this supercurrent-squared deformation in higher dimensions is related to the Dirac-Born-Infeld action. ??? This supersymmetric deformation extends naturally to higher dimensions, % where the leading operator involves a four derivative coupling. 
%where we conjecture that this deformation provides a natural higher-dimensional analogue of $T \Tb$. For special cases, this deformation gives the supersymmetric Dirac or Dirac-Born-Infeld action. 

\end{abstract}

\end{titlepage}

\tableofcontents

\section{Introduction} \label{intro}

% New refs:
% - David's papers are \cite{Aharony:2018ics} and \cite{Aharony:2018bad}
% - The competitor paper is \cite{Baggio:2018rpv}
% - The Italians paper with TT Maxwell in 2d is \cite{Conti:2018jho}
% - The instantons paper is \cite{TERASHIMA2000292}

It was first observed by Zamolodchikov \cite{zamolodchikovExpectationValueComposite2004} that the composite operator $\det ( T ) = T_{zz} T_{\bar{z} \bar{z}} - T_{z \bar{z}}^2$, often referred to as $T \Tb$, in a two-dimensional field theory is very special. Consider a family of Lagrangians obtained by solving the differential equation
\begin{align}
    \frac{\partial}{\partial t} \mathcal{L}^{(t)} = - \det \left(  T^{(t)} \right) ,
    \label{det-T-def}
\end{align}
where $T^{(t)}$ is the stress-energy tensor of the theory $\mathcal{L}^{(t)}$, and not of the original theory $\mathcal{L}^{(0)}$. The parameter $t$ has mass dimension $-2$. If $\mathcal{L}^{(0)}$ is conformal, it is natural to assign dimension $(-1,-1)$ to $t$. The flow equation~\C{det-T-def} defines a curve in the space of theories with two remarkable properties.

\begin{enumerate}
    \item The deformation preserves integrability: if one begins with a theory $\mathcal{L}^{(0)}$ which is integrable, in the sense that the theory has infinitely many local integrals of motion, then the deformed theory $\mathcal{L}^{(t)}$ at finite $t$ is also integrable.
    
    \item The deformation is ``solvable.'' By this, we mean that one can make precise statements about properties of the deformed theory $\mathcal{L}^{(t)}$ in terms of the un-deformed theory $\mathcal{L}^{(0)}$. If the theory is put on a cylinder of radius $R$, the energy levels of the deformed theory are related to those of the un-deformed theory by a differential equation of inviscid Burgers type.
\end{enumerate}

The two properties above are independent. If we begin with a theory $\mathcal{L}^{(0)}$ which is not integrable -- say, a single boson subject to a generic potential -- one can still write down differential equations describing the %finite-size 
spectrum of $\mathcal{L}^{(t)}$ in terms of the data of $\mathcal{L}^{(0)}$, even though neither theory is integrable \cite{bonelliBarDeformationsClosed2018,cavagliaBarTDeformed2D2016}. The ability to make exact statements about the spectrum of the deformed theory is particularly remarkable because $\det(T)$ is an irrelevant operator. The deformed theory is still very mysterious though there are strong indications it is not a conventional local quantum field theory with local operators~\cite{Giveon:2017nie}. 

% Although the usual $T \Tb$ deformation preserves integrability, it breaks manifest supersymmetry. 

Suppose one now begins with an undeformed Lagrangian which can be written in a manifestly supersymmetric way as an integral over superspace. If one deforms the physical Lagrangian with a finite $T \Tb$ deformation as in (\ref{det-T-def}), the resulting Lagrangian no longer respects the symmetries made manifest by the superspace construction. This does not mean that the quantum theory defined by the finite $T \Tb$ deformation is not supersymmetric. Indeed there are good reasons to believe that the theory is supersymmetric, but with non-canonical supersymmetry transformations. The argument goes as follows:\footnote{This argument was kindly described to us by David Kutasov.} supersymmetry in the undeformed theory implies a Bose-Fermi degeneracy between states with non-zero energy. The $T \Tb$ deformation has the property that it precisely preserves this degeneracy~\cite{Aharony:2018bad, Aharony:2018ics}. Therefore, at least this implication of supersymmetry should persist in the deformed theory.     

However, the control supersymmetry provides over quantum aspects of a theory is most easily seen in a formulation where supersymmetry is manifest. The purpose of this paper is to describe a modification of the $T \Tb$ deformation which preserves manifest supersymmetry. Our proposed generalization is written as a flow equation for the superspace Lagrangian, which is therefore automatically supersymmetric. We will see that the object replacing $\det ( T^{(t)} )$ on the right side of (\ref{det-T-def}) is constructed from a superfield which contains the supercurrent and stress tensor. We will colloquially refer to this deformation, which will be defined in section \ref{section:supercurrent-squared}, as ``supercurrent-squared.'' After integrating over superspace, this supercurrent-squared deformation reduces to the usual $T \Tb$ deformation, plus additional couplings needed to preserve supersymmetry. %which vanish on-shell.

The layout of this paper is as follows: in section \ref{section:tt_and_susy}, we define the supercurrent-squared deformation for two-dimensional theories with $(1,1)$ supersymmetry. We discuss its relationship with another supercurrent multiplet known as the $\mathcal{S}$-multiplet, and prove that this deformation is ``solvable'' in the sense of giving a differential equation for the finite-volume spectrum. In section \ref{section:(1,1)}, we write down the partial differential equation for the supercurrent-squared deformed Lagrangian in $(1,1)$ theories, beginning from either a free undeformed theory or one with an arbitrary superpotential. In section \ref{section:(0,1)}, we present the supercurrent-squared deformation for $(0,1)$ theories. We write down the differential equation for the deformed Lagrangian for a class of undeformed theories, and discuss the relationship between the $(1,1)$ and $(0,1)$ deformations. Details of the calculations are found in Appendices \ref{11_appendix} and \ref{01-appendix}.

There have been several proposals for generalizations of the $T \Tb$ deformation to quantum field theories in higher dimensions. In \cite{Taylor:2018xcy}, Taylor describes an operator $\mathcal{T}_D$ which is quadratic in the stress tensor and shares many properties of $T \Tb$, but which can be defined in arbitrary $D$ space-time dimensions. This deformation operator has leading dimension $2D$. Unlike $T \Tb$ in  $D=2$, this operator does not have an unambiguous quantum definition in $D>2$. Interestingly, adding this deformation to the free Maxwell Lagrangian in $D$ space-time dimensions gives a leading-order term
\begin{align}
    \mathcal{T}_D = \tensor{F}{_\mu^\sigma} F_{\nu \sigma} \tensor{F}{^\mu^\rho} \tensor{F}{^\nu_\rho} + \left( - \frac{1}{2} + \frac{D}{16} - \frac{\left(1 - \frac{D}{4} \right)^2}{D-1} \right) (F_{\alpha \beta} F^{\alpha \beta} )^2 ,
\end{align}
which matches the $F^4$ terms of the super-DBI action in $D=4$ but not in other dimensions.

Another interesting deformation of large $N$ CFTs by an operator built out of both the stress tensor and matter fields is presented in \cite{hartmanHolographyFiniteCutoff2018}. In this proposal, the deformation operator again has leading dimension $2D$. An Appendix of \cite{cardyOverlineDeformationQuantum2018} discusses obstructions to generalizing the interpretation of $T \Tb$ as random geometry to higher dimensions, concluding that the deformation is no longer topological in $D>2$. In \cite{bonelliBarDeformationsClosed2018}, the authors discuss both the generalization to $\left| \det T \right|^{1/\alpha}$, with $\alpha = D - 1$, as well as an alternate proposal which is also quadratic in the stress tensor and reduces to the Burgers' equation. Lastly, an intriguing connection between the $T \Tb$ deformation and Jackiw–Teitelboim gravity is discussed in \cite{dubovskyBarTPartitionFunction2018,Dubovsky2017}, where the authors speculate that a higher-dimensional version of this procedure might explain the $M_{\text{weak}} \sim \sqrt{M_{\text{Pl}} \,  E_{\text{vac}}}$ coincidence.

However, our work suggests a different generalization of $T \Tb$ to higher dimensions. We are motivated both by string theory and by the beautiful connection between the $T \Tb$ deformation of free scalars in $D=2$ and the Dirac action explained in~\cite{cavagliaBarTDeformed2D2016, bonelliBarDeformationsClosed2018}. For other interesting connections between the $T \Tb$ deformation and string theory, see~\cite{Caselle:2013dra, Baggio:2018gct, Dei:2018mfl}.  A supersymmetric theory in any number of space-time dimensions has conserved supercurrents associated with its supercharges. We conjecture that a deformation built from bilinears in these supercurrents might be ``solvable,'' in the sense discussed above, at least for models with a sufficient degree of supersymmetry. 

String theory branes provide some motivation for this conjecture for theories with extended supersymmetry. The Dirac-Born-Infeld (DBI) action describes the slowly varying dynamics of a single Dp-brane of string theory. Ignoring the string dilaton and assuming a flat space-time metric, the DBI action takes the form, 
\be\label{DBI}
S_{DBI} = - T_p \int d^{p+1}\s \, \sqrt{- \det{\left(\eta_{\mu\nu} + \partial_\mu \phi \partial_\nu \phi + \alpha F_{\mu\nu}\right)}}.
\ee
There are two dimensionful parameters $T_p$ and $\alpha$; $F_{\mu\nu}$ is an abelian field strength. 
Usually the appearance of the Dirac action is tied to spontaneously broken Poincar\'e invariance. However, the DBI action can also be motivated differently in models with maximal supersymmetry by asking: what irrelevant deformations of the low-energy brane theory are compatible with maximal supersymmetry? While there is no complete proof that the DBI action is the unique answer to that question, in an approximation where acceleration terms are neglected, there is considerable evidence that this should be the case. 

Certainly the leading irrelevant operators in the expansion of~\C{DBI}\ are determined in models with the maximal sixteen supercharges~\cite{Tseytlin:1999dj, Bergshoeff:1986jm, Metsaev:1987by, Metsaev:1987qp, Paban:1998ea, Paban:1998qy, Lin:2015ixa, Chen:2015hpa, Garousi:2017fbe, Heydeman:2017yww}. For models with less supersymmetry, Cecotti and Ferrara pointed out long ago that when one expands the four-dimensional $\mathcal{N} = 1$ super-DBI action to first order in field strengths, the leading term is simply the square of the supercurrent multiplet~\cite{cecottiSupersymmetricBorninfeldLagrangians1987}. This suggests that deforming with supercurrent-squared, or a close cousin of this operator, might lead to the Dirac or DBI actions in higher dimensions. The deformation can then also be studied for the case of interacting theories. It is important to note that the $T \Tb$ deformation of the free Maxwell theory in $D=2$ surprisingly does not lead to the Born-Infeld action~\cite{Conti:2018jho}.  However, the authors of \cite{Conti:2018jho} discuss an interesting different connection between two-dimensional $T \Tb$ deformations and the Born-Infeld theory in $D=4$, which might well be related to our conjecture involving the supercurrent-squared operator.

%as it is in two dimensions. This deformation is lower-order in a derivative expansion than $\det ( T )$: the leading term of supercurrent-squared will have four derivatives in any space-time dimension, while the first-order piece of the $\det ( T )$ deformation contains a term with $2^D$ derivatives in $D$ space-time dimensions.

\vskip 0.1in
\noindent {\bf Note Added:} while this paper was in the final stages of editing, an interesting paper appeared with overlapping observations~\cite{Baggio:2018rpv}.

\section{\texorpdfstring{$T \Tb$}{Lg} and Supersymmetry}\label{section:tt_and_susy}
% In this section we discuss our proposal of supercurrent square in superspace and why they are integrable.
In this section we propose a solvable deformation that is compatible with supersymmetry. As we will discuss, the remarkable property of the $T\bar T$ deformation follows from continuity equations. In the supersymmetric case, we  describe analogous relations based on the conservation laws in superspace.

%so we propose an analogue based on the conservation laws in superspace.

\subsection{Bi-spinor conventions}
To fix conventions, we consider two-dimensional field theories in Lorentzian signature with coordinates $(x^0, x^1)$. It will be convenient to change coordinates to light-cone variables using bi-spinor notation. That is, we define
\begin{align}
    x^{\pm \pm} = \frac{1}{\sqrt{2}} \left( x^0 \pm x^1 \right) , 
\end{align}
and write the corresponding derivatives as $\partial_{\pm \pm} = \frac{1}{\sqrt{2}}(\partial_0 \pm \partial_1)$. In these conventions, we have $\partial_{\pm \pm} x^{\pm \pm} = 1$ and $\partial_{\pm \pm} x^{\mp \mp} = 0$.

Spinors in two dimensions carry a single index which is raised or lowered as follows:
\begin{align}
    \psi^+ = - \psi_-, \qquad   \psi^- = \psi_+ . 
\end{align}
The advantage of writing all vector indices as pairs of spinor indices is that it allows us to more easily compare terms in equations which involve a combination of spinor, vector, spinor-vector, and tensor quantities. For instance, in this notation the supercurrent has components $S_{+++}, S_{---}, S_{+--}$, and $S_{-++}$, which we can immediately identify as a spinor-vector because it has three indices. Likewise, the stress-energy tensor carries two vector indices so its components will have four bispinor indices; they are written as $T_{++++}, T_{----}, T_{++--} = T_{--++}$.

When we consider $(1,1)$ supersymmetric theories, we will introduce anticommuting coordinates $\theta^{\pm}$. The corresponding supercovariant derivatives are defined in our conventions as
\begin{align}
    D_{\pm} = \frac{\partial}{\partial \theta^{\pm}} + \theta^{\pm} \partial_{\pm \pm} ,
    \label{D-def}
\end{align}
which satisfy $D_{\pm} D_{\pm} = \partial_{\pm \pm}$ and $\left\{ D_+ , D_- \right\} = 0$. There are also two supercharges $Q_{\pm}$ given by
\begin{align}
    Q_{\pm} = \frac{\partial}{\partial \theta^{\pm}} - \theta^{\pm} \partial_{\pm \pm},
\end{align}
which satisfy $Q_{\pm} Q_{\pm} = - \partial_{\pm \pm}$.

\subsection{\texorpdfstring{Review of $T \Tb$}{Lg}}\label{TTb-review}
% Review the basic of TT bar deformation and use free theory as an example.

Let us review some basics of the $T \Tb$ deformation with no supersymmetry involved. An integrable field theory contains an infinite set of local integrals of motion generated by conserved currents. Among those currents, we are mostly interested in the stress energy tensor, $(T_\wwww, T_{\pm\pm\mp\mp})$, which satisfies the continuity equations
\begin{align}
    &\partial_\mm T_\pppp + \partial_\pp T_\ppmm = 0 \nonumber \\
    &\partial_\pp T_\mmmm + \partial_\mm T_\mmpp = 0 .
\end{align}
Given this set of continuity equations, the authors of \cite{smirnovSpaceIntegrableQuantum2017} suggest considering spinless composite operators by first constructing bilinear operators, $T_\pppp(x)T_\mmmm(\xp)$ and\newline $T_\ppmm(x)T_\mmpp(\xp)$, and taking the limit $x\rightarrow\xp$. Although such limit is usually singular, the combination
\begin{align}
    T_\pppp(x)T_\mmmm(\xp) - T_\ppmm(x)T_\mmpp(\xp)
\end{align}
can be shown to contain no non-derivative divergences in its OPE. It is then natural to define a local operator $T\bar T (x)$ by
\begin{align}
    T\bar T(x) + \ldots = T_\pppp(x)T_\mmmm(\xp) - T_\ppmm(x)T_\mmpp(\xp)
\end{align}
where $\ldots$ contains derivative terms that are not necessarily regular. A $T\bar T$ deformed theory with action $S^{(t)}$ is a two dimensional integrable field theory deformed by the local operator $T\bar T(x)$ that satisfies the flow equation
\begin{equation}\label{equation:flow}
    \frac{\partial}{\partial t} S^{(t)} = -\int d^2x\ T\bar T(x) .
\end{equation}
The undeformed theory is the initial condition at $t=0$. It should be emphasized that $T\bar T$ in (\ref{equation:flow}) is constructed from the stress energy tensor of the deformed action $S^{(t)}$, thus it depends on the coupling $t$ implicitly and the flow equation becomes a non-linear differential equation. The significance of this deformation is it preserves the integrability. Furthermore, when the spatial direction is compactified on a circle of circumference $R$, the energy spectrum of the deformed theory is determined by the inviscid Burgers' equation,
\begin{align}
    \frac{\partial}{\partial t} E_n(t,R) = E_n(t,R)\frac{\partial}{\partial R}E_n(t,R) + \frac{P^2_n}{R}.
\end{align}
Here $E_n(t,R)$ is the energy level and $P(R)$ is the momentum of that state.

As a simple example, consider a free massless scalar $\phi$ in two dimensions with action given by
\begin{align}
    S^{(0)} = \int d^2x\ \partial_\pp\phi\partial_\mm\phi 
\end{align}
Here $d^2x$ means the measure $dx^1 dx^2$. Solving the differential equation (\ref{equation:flow}), the $T\bar T$ deformed action turns out to be~\cite{krausCutoffAdSBarT2018}
\begin{align}\label{equation:Nambu-Goto}
    S^{(t)} = \int d^2x\ \frac{1}{2t} \lrb -1 + \sqrt{1 + 4t
    \partial_\pp\phi\partial_\mm\phi} \rrb.
\end{align}
It was noted in \cite{cavagliaBarTDeformed2D2016} that this is the Nambu-Goto action in static gauge. 

\subsection{Supercurrent-squared}\label{section:supercurrent-squared}
% Derive the conservation law in superspace and propose supercurrent square as a candidate of supersymmetrized TT bar deformation.

Because the usual $T \Tb$ deformation discussed in section (\ref{TTb-review}) is built from the Noether current for spatial translations, we will generalize this construction by writing a manifestly supersymmetric Noether current associated with translations in superspace. For concreteness, we will work in the $(1,1)$ theory, but a similar calculation in $(0,1)$ will be described in section \ref{section:(0,1)}.

Consider a supersymmetric Lagrangian which is written as an integral over $(1,1)$ superspace as $\mathcal{L} = \int d^2 \theta \, \mathcal{A}$. We allow $\mathcal{A}$ to depend on a superfield $\Phi$ and a particular set of $\Phi$ derivatives listed below:
\be \mathcal{A} = \mathcal{A}\left( \Phi, D_+ \Phi, D_- \Phi, \partial_{++} \Phi, \partial_{--} \Phi, D_+ D_- \Phi \right).
\ee 
The supercovariant derivatives $D_{\pm}$ are defined in (\ref{D-def}). The superspace equation of motion associated with this Lagrangian is
\begin{align}
	\frac{\delta \mathcal{A}}{\delta \Phi} & =  D_+ \left( \frac{\delta \mathcal{A}}{\delta D_+ \Phi} \right) + D_- \left( \frac{\delta \mathcal{A}}{\delta D_- \Phi} \right) + \partial_{++} \left( \frac{\delta \mathcal{A}}{\delta \partial_{++} \Phi} \right) \cr & + \partial_{--} \left( \frac{\delta \mathcal{A}}{\delta \partial_{--} \Phi} \right) - D_+ D_- \left( \frac{\delta \mathcal{A}}{\delta D_+ D_- \Phi} \right) . 
\end{align}
As in the derivation of the usual stress tensor $T$, we now consider a spatial translation of the form $\delta x^{\pm \pm} = a^{\pm \pm}$ for some constant $a^{\pm \pm}$. The variation $\delta \mathcal{A}$ of the superspace Lagrangian is given by
\begin{equation}\label{general_superspace_variation} 
\begin{split}
%\begin{align}
	\delta \mathcal{A} &= D_+ \left( \delta \Phi \frac{\delta \mathcal{A}}{\delta D_+ \Phi} \right) + D_- \left( \delta \Phi \frac{\delta \mathcal{A}}{\delta D_- \Phi} \right) + \partial_{++} \left( \delta \Phi \frac{\delta \mathcal{A}}{\delta \partial_{++} \Phi} \right) \cr
    &+ \partial_{--} \left( \delta \Phi \frac{\delta \mathcal{A}}{\delta \partial_{--} \Phi} \right) + \frac{1}{2} \left( D_+ \left( \frac{\delta \mathcal{A}}{\delta D_+ D_- \Phi} D_- \delta \Phi \right) + D_- \left( \delta \Phi D_+ \frac{\delta \mathcal{A}}{\delta D_+ D_- \Phi} \right) \right) \cr
    &- \frac{1}{2} \left( D_- \left( \frac{\delta \mathcal{A}}{\delta D_+ D_- \Phi} D_+ \delta \Phi \right) + D_+ \left( \delta \Phi D_- \frac{\delta \mathcal{A}}{\delta D_+ D_- \Phi} \right)  \right) \cr
    &- \delta \Phi \left( - \frac{\delta \mathcal{A}}{\delta \Phi} + D_+ \frac{\delta \mathcal{A}}{\delta D_+ \Phi} + D_- \frac{\delta \mathcal{A}}{\delta D_- \Phi} + \partial_{++} \frac{\delta \mathcal{A}}{\delta \partial_{++} \Phi} + \partial_{--} \frac{\delta \mathcal{A}}{\delta \partial_{--} \Phi}  \right. \cr & \left. - D_+ D_- \frac{\delta \mathcal{A}}{\delta D_+ D_- \Phi} \right) . 
%\end{align}
\end{split} 
\end{equation}
Here we have chosen to symmetrize the term involving $D_+ D_- \frac{\delta \mathcal{A}}{\delta D_+ D_- \Phi}$ using $\left\{ D_+ , D_- \right\} = 0$.

The last two lines of (\ref{general_superspace_variation}) are the superspace equation of motion; we now specialize to the case of on-shell variations, for which this last term vanishes. Further, the left side of (\ref{general_superspace_variation}) is $\delta \mathcal{A} = a^{++} \partial_{++} \mathcal{A} + a^{--} \partial_{--} \mathcal{A}$, which is a total derivative. We use $\partial_{\pm \pm} = D_{\pm} D_{\pm}$ to express (\ref{general_superspace_variation}) in the form
\begin{align}
\begin{split}
	0 &= a^{++} D_+ \Big[ \partial_{++} \Phi \frac{\delta \mathcal{A}}{\delta D_+ \Phi} + D_+ \left(  \partial_{++} \Phi \frac{\delta \mathcal{A}}{\delta \partial_{++} \Phi} \right) + \frac{1}{2} \frac{\delta \mathcal{A}}{\delta D_+ D_- \Phi} D_- \left( \partial_{++} \Phi  \right) \label{superspace-T-conservation} \\
	&\qquad \qquad \qquad - \frac{1}{2} \partial_{++} \Phi D_- \left( \frac{\delta \mathcal{A}}{\delta D_+ D_- \Phi} \right) - D_+ \mathcal{A} \Big]  \\
    &+ a^{++} D_- \Big[ \partial_{++} \Phi \frac{\delta \mathcal{A}}{\delta D_- \Phi} + D_- \left(  \partial_{++} \Phi  \frac{\delta \mathcal{A}}{\delta \partial_{--} \Phi} \right)  - \frac{1}{2} \frac{\delta \mathcal{A}}{\delta D_+ D_- \Phi} D_+ \left( \partial_{++} \Phi \right)  \\
    &\qquad \qquad \qquad + \frac{1}{2} \partial_{++} \Phi D_+ \left( \frac{\delta \mathcal{A}}{\delta D_+ D_- \Phi} \right) \Big]  \\
	&+ a^{--} D_+ \Big[ \partial_{--} \Phi \frac{\delta \mathcal{A}}{\delta D_+ \Phi} + D_+ \left( \partial_{--} \Phi \frac{\delta \mathcal{A}}{\delta \partial_{++} \Phi} \right) + \frac{1}{2} \frac{\delta \mathcal{A}}{\delta D_+ D_- \Phi} D_- \left( \partial_{--} \Phi \right)  \\
	&\qquad \qquad \qquad - \frac{1}{2} \partial_{--} \Phi D_- \left( \frac{\delta \mathcal{A}}{\delta D_+ D_- \Phi} \right) \Big]  \\
    &+ a^{--} D_- \Big[  \partial_{--} \Phi \frac{\delta \mathcal{A}}{\delta D_- \Phi} + D_- \left( \partial_{--} \Phi \frac{\delta \mathcal{A}}{\delta \partial_{--} \Phi} \right) - \frac{1}{2} \frac{\delta \mathcal{A}}{\delta D_+ D_- \Phi} D_+ \left( \partial_{--} \Phi \right)  \\
    &\qquad \qquad \qquad + \frac{1}{2} \partial_{--} \Phi D_+ \left( \frac{\delta \mathcal{A}}{\delta D_+ D_- \Phi} \right) - D_- \mathcal{A} \Big] .
\end{split}
\end{align}
This equation gives a conservation law for a superfield $\mathcal{T}$ which we define by
\begin{align}
	\mathcal{T}_{++-} &= \partial_{++} \Phi \frac{\delta \mathcal{A}}{\delta D_+ \Phi} + D_+ \left(  \partial_{++} \Phi \frac{\delta \mathcal{A}}{\delta \partial_{++} \Phi} \right) + \frac{1}{2} \frac{\delta \mathcal{A}}{\delta D_+ D_- \Phi} D_- \left( \partial_{++} \Phi  \right)  \nonumber \\
	&\quad - \frac{1}{2} \partial_{++} \Phi D_- \left( \frac{\delta \mathcal{A}}{\delta D_+ D_- \Phi} \right) - D_+ \mathcal{A}  , \nonumber \\
    \mathcal{T}_{+++} &= \partial_{++} \Phi \frac{\delta \mathcal{A}}{\delta D_- \Phi} + D_- \left(  \partial_{++} \Phi  \frac{\delta \mathcal{A}}{\delta \partial_{--} \Phi} \right)  - \frac{1}{2} \frac{\delta \mathcal{A}}{\delta D_+ D_- \Phi} D_+ \left( \partial_{++} \Phi \right) \nonumber \\
    &\quad + \frac{1}{2} \partial_{++} \Phi D_+ \left( \frac{\delta \mathcal{A}}{\delta D_+ D_- \Phi} \right) , \nonumber \\
    \mathcal{T}_{---} &= \partial_{--} \Phi \frac{\delta \mathcal{A}}{\delta D_+ \Phi} + D_+ \left( \partial_{--} \Phi \frac{\delta \mathcal{A}}{\delta \partial_{++} \Phi} \right) + \frac{1}{2} \frac{\delta \mathcal{A}}{\delta D_+ D_- \Phi} D_- \left( \partial_{--} \Phi \right) \label{final_ttbar_general} \\
    &\quad - \frac{1}{2} \partial_{--} \Phi D_- \left( \frac{\delta \mathcal{A}}{\delta D_+ D_- \Phi} \right) , \nonumber \\
    \mathcal{T}_{--+} &=  \partial_{--} \Phi \frac{\delta \mathcal{A}}{\delta D_- \Phi} + D_- \left( \partial_{--} \Phi \frac{\delta \mathcal{A}}{\delta \partial_{--} \Phi} \right) - \frac{1}{2} \frac{\delta \mathcal{A}}{\delta D_+ D_- \Phi} D_+ \left( \partial_{--} \Phi \right) \nonumber \\
    &\quad + \frac{1}{2} \partial_{--} \Phi D_+ \left( \frac{\delta \mathcal{A}}{\delta D_+ D_- \Phi} \right) - D_- \mathcal{A} \nonumber . 
\end{align}
In terms of $\mathcal{T}$, then, equation (\ref{superspace-T-conservation}) implies the superspace conservation laws:
\begin{align}\label{equation:conservation in superspace}
    D_+ \mathcal{T}_{++-} + D_- \mathcal{T}_{+++} &= 0 , & D_+ \mathcal{T}_{---} + D_- \mathcal{T}_{--+} &= 0 .
\end{align}
We are now in a position to propose the supercurrent-squared deformation. Consider a one-parameter family of superspace Lagrangians labeled by $t$, which satisfy the ordinary differential equation
\begin{align}
	\frac{\partial}{\partial t} \mathcal{A}^{(t)} = \mathcal{T}_{+++}^{(t)} \mathcal{T}_{---}^{(t)} - \mathcal{T}_{--+}^{(t)} \mathcal{T}_{++-}^{(t)} ,
    \label{ttbar_general_flow}
\end{align}
where $\mathcal{T}^{(t)}$ is the supercurrent superfield (\ref{final_ttbar_general}) computed from the superspace Lagrangian $\mathcal{A}^{(t)}$. This uniquely defines the supercurrent-squared deformation of an initial Lagrangian $\mathcal{A}^{(0)}$ at finite deformation parameter $t$.

\subsection{Reduction to components for a free theory}\label{section:free_components}

To illutrate the relationship between the flow equation (\ref{ttbar_general_flow}) and the usual $T \Tb$ operator, let us explicitly compute the components of the supercurrent-squared deformation for a free $(1,1)$ superspace Lagrangian
\begin{align}
    \mathcal{A} = D_+ \Phi D_- \Phi ,
\end{align}
where $\Phi$ is a superfield with component expansion
\begin{align}
    \Phi = \phi + i \theta^+ \psi_+ + i \theta^- \psi_- + \theta^+ \theta^- f .
\end{align}
The entries of $\mathcal{T}$, defined by (\ref{final_ttbar_general}), for the free theory are
\begin{align}
\begin{split}
    \mathcal{T}_{++-} &= \partial_{++} \Phi D_- \Phi - D_+ \left( D_+ \Phi D_- \Phi \right) , \label{free_superspace} \\
    \mathcal{T}_{+++} &= - \partial_{++} \Phi D_+ \Phi ,  \\
    \mathcal{T}_{---} &= \partial_{--} \Phi D_- \Phi ,  \\
    \mathcal{T}_{--+} &= - \partial_{--} \Phi D_+ \Phi - D_- \left( D_+ \Phi D_- \Phi \right)  .
\end{split}
\end{align}
In components, (\ref{free_superspace}) is
\begin{align}
    \mathcal{T}_{++-} =& - i \psi_+ f + \theta^+ \left( - f \partial_{++} \phi + \psi_+ \partial_{++} \psi_- \right) + \theta^- \left( - f^2 - \psi_+ \partial_{--} \psi_+ \right) \nonumber \\
    & + i \theta^+ \theta^- \left( - \partial_{++} \phi \partial_{--} \psi_+ - \partial_{++} \psi_+ \partial_{--} \phi - f \partial_{++} \psi_- + \psi_- \partial_{++} f + \partial_{++} \left( \psi_+ \partial_{--} \phi - \psi_- f \right) \right) , \nonumber \\
    \mathcal{T}_{+++} =& - i \psi_+ \partial_{++} \phi - \theta^+ \left( \psi_+ \partial_{++} \psi_+ + \left( \partial_{++} \phi \right)^2 \right) - \theta^- \left( f \partial_{++} \phi + \psi_+ \partial_{++} \psi_- \right) \nonumber \\
    & - i \theta^+ \theta^- \left( 2 \partial_{++} \phi \partial_{++} \psi_- + \psi_+ \partial_{++} f - f \partial_{++} \psi_+ \right) , \nonumber \\
    \mathcal{T}_{---} =&  \, i \psi_- \partial_{--} \phi + \theta^+ \left( \psi_- \partial_{--} \psi_+ - f \partial_{--} \phi \right) + \theta^- \left( \psi_- \partial_{--} \psi_- + \left( \partial_{--} \phi \right)^2 \right) \nonumber \\
    & + i \theta^+ \theta^- \left( \psi_- \partial_{--} f - f \partial_{--} \psi_- - 2 \partial_{--} \phi \partial_{--} \psi_+ \right) , \label{free_superspace_components}  \\
    \mathcal{T}_{--+} =& - i\psi_- f + \theta^+ \left( f^2 + \psi_- \partial_{++} \psi_- \right) + \theta^- \left( -f \partial_{--} \phi - \psi_- \partial_{--} \psi_+ \right) \nonumber \\
    & + i \theta^+ \theta^- \left( -\partial_{--} \phi \partial_{++} \psi_- + f \partial_{--} \psi_+ - \partial_{--} \psi_- \partial_{++} \phi - \psi_+ \partial_{--} f + \partial_{--} \left( \psi_+ f + \psi_- \partial_{++} \phi \right) \right). \nonumber
\end{align}
To compare with the bosonic $T \Tb$ deformation, we identify the components of the usual stress tensor $T$ for the theory of a free boson $\phi$ and fermions $\psi_{\pm}$ which one obtains by performing the integrals over $\theta^{\pm}$. In our conventions, these take the form:
\begin{align}
\begin{split}
    T_{++++} &= \left( \partial_{++} \phi \right)^2 + \psi_+ \partial_{++} \psi_+ , \\
    T_{----} &= \left( \partial_{--} \phi \right)^2 + \psi_- \partial_{--} \psi_- .
\end{split}
\end{align}
We will also drop terms involving the auxiliary field $f$, since in the bosonic part of the supercurrent-squared deformation, these terms vanish after integrating out $f$ using its equation of motion. Then the bilinears appearing in our flow equation (\ref{ttbar_general_flow}) are
\begin{align}
\begin{split}
    \mathcal{T}_{+++} \mathcal{T}_{---} &= \psi_+ \psi_- \partial_{++} \phi \partial_{--} \phi + i \theta^+ \left( \psi_+ \psi_- \partial_{++} \phi \partial_{--} \psi_+ - T_{++++} \psi_- \partial_{--} \phi \right) \label{sc-square-no-f} \\
    &\quad + i \theta^- \left( \psi_+ \partial_{++} \phi T_{----} + \psi_+ \psi_- \partial_{++} \psi_- \partial_{--} \phi \right) - \theta^+ \theta^- \big( T_{++++} T_{----} \\
    &\quad + 2 \partial_{++} \phi \partial_{--} \phi \left( \psi_+ \partial_{--} \psi_+ +  \psi_- \partial_{++} \psi_- \right) - \psi_- \partial_{++} \psi_- \psi_+ \partial_{--} \psi_+ \big) , \\
    \mathcal{T}_{++-} \mathcal{T}_{--+} &= - 2 \theta^+ \theta^- \left( \psi_+ \partial_{--} \psi_+ \psi_- \partial_{++} \psi_- \right) .
\end{split}
\end{align}
The superspace integral of the deformation $\mathcal{T}_{+++} \mathcal{T}_{---} + \mathcal{T}_{++-} \mathcal{T}_{--+}$ picks out the top component, which is
\begin{align}
\begin{split}
    \int &d^2 \theta \, \left( \mathcal{T}_{+++} \mathcal{T}_{---} + \mathcal{T}_{++-} \mathcal{T}_{--+} \right) = \label{sc-square-integrated} \\
    &- T_{++++} T_{----} - 2 \partial_{++} \phi \partial_{--} \phi \left( \psi_+ \partial_{--} \psi_+ + \psi_- \partial_{++} \psi_- \right) - \psi_- \partial_{++} \psi_- \psi_+ \partial_{--} \psi_+ .
\end{split}
\end{align}
We see that (\ref{sc-square-integrated}) contains the usual $T \Tb$ deformation, given in our bi-spinor notation by $-T_{++++} T_{----}$, along with extra terms which are all proportional to the fermion equations of motion, $\partial_{\pm \pm} \psi_{\mp} = 0$. These added terms vanish on-shell and, as we will argue in section (\ref{section:solvable}), do not spoil solvability: the energy levels of the deformed theory can still be expressed in terms of those in the undeformed theory, as in the purely bosonic $T \Tb$ case.

\subsection{Relationship with the $\mathcal{S}$-multiplet}

The $(1,1)$ superfield $\mathcal{T}$ contains the conserved stress-energy tensor $T_{\mu \nu}$ and the supercurrent $S_{\mu \alpha}$. Such current multiplets have received much attention in the literature; the first construction for four-dimensional theories was the FZ multiplet \cite{Ferrara:1974pz}, which was later shown to be a special case of the more general $\mathcal{S}$-multiplet \cite{dumitrescuSupercurrentsBraneCurrents2011a}.

For the two-dimensional theories we consider here, it is known that the $\mathcal{S}$-multiplet is the most general multiplet containing the stress tensor and supercurrent, subject to assumptions that the multiplet be indecomposable and contain no other operators with spin greater than one. Since our supercurrent superfield $\mathcal{T}$ satisfies these properties, it must be equivalent to the $\mathcal{S}$-multiplet. As we will show, the four superfields contained in $\mathcal{T}$ are identical to the four superfields of the $\mathcal{S}$-multiplet, up to terms which vanish on-shell and therefore do not affect the conservation equations for the currents.

The $\mathcal{S}$-multiplet is a reducible but indecomposable set of two superfields $\mathcal{S}$ and $\chi$ satisfying the constraints
\begin{align}
\begin{split}
    D_{\mp} \mathcal{S}_{\pm \pm \pm} &= D_{\pm} \chi_{\pm} , \label{s-multiplet-constraints} \\
    D_- \chi_+ &= D_+ \chi_- . 
\end{split}
\end{align}
In components, the $\mathcal{S}$-multiplet for $(1,1)$ theories contains the usual stress tensor $T_{\mu \nu}$, the supercurrent $S_{\mu \alpha}$, and a vector $Z_\mu$ which is associated with a scalar central charge:
\begin{align}
\begin{split}
    \mathcal{S}_{+++} &= S_{+++} + \theta^+ T_{++++} + \theta^- Z_{++} - \theta^+ \theta^- \partial_{++} S_{-++} , \label{s-multiplet} \\
    \mathcal{S}_{---} &= S_{---} + \theta^+ Z_{--} + \theta^- T_{----} + \theta^+ \theta^- \partial_{--} S_{+--} , \\
    \chi_+ &= S_{-++} + \theta^+ Z_{++} + \theta^- T_{++--} - \theta^+ \theta^- \partial_{++} S_{+--} , \\
    \chi_- &= S_{+--} + \theta^+ T_{++--} + \theta^- Z_{--} + \theta^+ \theta^- \partial_{--} S_{-++} . 
\end{split}
\end{align}
In terms of these component fields, the constraints (\ref{s-multiplet-constraints}) give conservation equations for the currents:
\begin{align}
\begin{split}
    \partial_{++} T_{----} + \partial_{--} T_{++--} &= 0 = \partial_{++} T_{--++} + \partial_{--} T_{++++} , \\
    \partial_{++} S_{+--} + \partial_{--} S_{+++} &= 0 = \partial_{++} S_{---} + \partial_{--} S_{-++} , \\
    \partial_{++} Z_{--} + \partial_{--} Z_{++} &= 0 .
\end{split}
\end{align}

We claim that the components (\ref{free_superspace_components}) of our superspace supercurrent are the same as those in the two superfields $\mathcal{S}$ and $\chi$ appearing in the $(1,1)$ $\mathcal{S}$-multiplet (\ref{s-multiplet}), up to signs and terms which vanish on-shell. In particular, after discarding terms which are proportional to the equations of motion, we find the identifications: \begin{align}
\mathcal{S}_{\pm \pm \pm} = \mp \mathcal{T}_{\pm \pm \pm}, \qquad \chi_+ = \mathcal{T}_{++-}, \qquad \chi_- = \mathcal{T}_{--+}. 
\end{align}

We will check this explicitly for the free theory, $\mathcal{A} = D_+ \Phi D_- \Phi$, for which we computed the components of $\mathcal{T}$ in section (\ref{section:free_components}). Writing only those terms that survive when the component equations of motion $f=0$, $\partial_{++} \psi_- = 0 = \partial_{--} \psi_+$, and $\partial_{++} \partial_{--} \phi = 0$ are all satisfied, (\ref{free_superspace_components}) becomes
\begin{align}
\begin{split}
    \mathcal{T}_{++-} &\overset{\text{on-shell}}{=} 0 , \label{free_superspace_on_shell} \\
    \mathcal{T}_{+++} &\overset{\text{on-shell}}{=} - i \psi_+ \partial_{++} \phi - \theta^+ \left( \psi_+ \partial_{++} \psi_+ + \left( \partial_{++} \phi \right)^2 \right) , \\
    \mathcal{T}_{---} &\overset{\text{on-shell}}{=} i \psi_- \partial_{--} \phi + \theta^- \left( \psi_- \partial_{--} \psi_- + \left( \partial_{--} \phi \right)^2 \right) , \\
    \mathcal{T}_{--+} &\overset{\text{on-shell}}{=} 0 .
\end{split}
\end{align}
For the free $(1,1)$ superfield considered here, the supercurrent is given in our conventions by
\begin{align}
\begin{split}
    S_{+++} &= \psi_+ \partial_{++} \phi , \\
    S_{---} &= \psi_- \partial_{--} \phi , \\
    S_{+--} &= 0 = S_{-++} ,
\end{split}
\end{align}
while the stress tensor components are as in (\ref{bos-T-free}). To find expressions for the scalar central charge current $Z_{\pm \pm}$, we use the supersymmetry algebra implied by the $\mathcal{S}$-multiplet constraints, which gives
\begin{align}
\begin{split}
    \left\{ Q_{\pm} , S_{\pm \pm \pm} \right\} &= T_{\pm \pm \pm \pm} , \label{S-algebra} \\
    \left\{ Q_{\pm} , S_{\pm \mp \mp} \right\} &= T_{\pm \pm \mp \mp} ,  \\
    \left\{ Q_{\pm} , S_{\mp \pm \pm} \right\} &= Z_{\pm \pm} ,  \\
    \left\{ Q_{\pm} , S_{\mp \mp \mp} \right\} &= Z_{\mp \mp} . 
\end{split}
\end{align}
Note that the $\mathcal{S}$-multiplet constraints only hold when the conservation equations for the currents hold, so the relations (\ref{S-algebra}) should be viewed as an on-shell algebra. Acting with the supercharges $Q_{\pm}$ on the stress tensor and supercurrent components, one finds that $Z_{--} \sim \psi_- \partial_{--} \psi_+$ and $Z_{++} \sim \psi_+ \partial_{++} \psi_-$, both of which vanish when the fermion equations of motion are satisfied.

Thus, after imposing the equations of motion, we can write our supercurrent superfield components as
\begin{align}
    \mathcal{T}_{++-} &= \chi_+ = 0, & \mathcal{T}_{--+} &= 0 = \chi_{-}, \nonumber \\
    \mathcal{T}_{+++} &= - S_{+++} - \theta^+ T_{++++} = - \mathcal{S}_{+++}, & \mathcal{T}_{---} &= S_{---} + \theta^- T_{----} = \mathcal{S}_{---} .
\end{align}
Since terms which vanish on-shell do not affect conservation equations, one can view $\mathcal{T}$ as an improvement transformation of the $\mathcal{S}$-multiplet. The constraint equation $D_{\mp} \mathcal{S}_{\pm \pm \pm} - D_{\pm} \chi_{\pm} = 0$ is expressed by our conservation equations $D_+ \mathcal{T}_{++-} + D_- \mathcal{T}_{+++} = 0$ and $D_+ \mathcal{T}_{---} + D_- \mathcal{T}_{+--} = 0$.

\subsection{Why are superspace deformations solvable?}\label{section:solvable}
% Argue the above deformation is solvable.
In this section we prove the theory deformed by (\ref{ttbar_general_flow}) is solvable just like the usual $T\bar T$ deformation. Let's begin with the conservation law in superspace (\ref{equation:conservation in superspace}). It is straightforward to solve these constraints in components by using the conservation of the stress energy tensor:
\begin{align}
\begin{split}
    &\T_\ppp = H_\ppp - \tp T_\pppp - \tm W_\pp + \tp\tm G_\ppp, \\
    &\T_\mmm = H_\mmm + \tm T_\mmmm + \tp W_\mm - \tp\tm G_\mmm, \\
    &\T_\mmp = H_\mmp - \tp T_\mmpp - \tm W_\mm + \tp\tm G_\mmp, \\
    &\T_\ppm = H_\ppm + \tm T_\ppmm + \tp W_\pp - \tp\tm G_\ppm .
\end{split}
\end{align}
Here $(H_\www,H_{\mp\mp\pm})$ denote the lowest components of $\T$ while $(G_\www,G_{\mp\mp\pm})$ are its highest components. The conservation law in (\ref{equation:conservation in superspace}) implies constraints on $G$ and $H$:
\begin{align}\label{equation:Conservation G H}
\begin{split}
    &G_{\mp\mp\pm} = \partial_{\pm\pm} H_{\mp\mp\mp} , \\
    &G_{\pm\pm\pm} = \partial_{\pm\pm} H_{\mp\pm\pm} .
\end{split}
\end{align}
In terms of these components, the deformation in (\ref{ttbar_general_flow}) becomes
\begin{align}
    \frac{\partial}{\partial t} \mathcal{L}^{(t)} =
    & -\int d^2\theta \lrb \T_\ppp\T_\mmm + \T_\ppm\T_\mmp \rrb \nonumber \\
    = & -\lrb T_\pppp T_\mmmm - T_\ppmm T_\mmpp \rrb \\
    & + \lrb H_\ppp G_\mmm - G_\ppp H_\mmm - H_\ppm G_\mmp + G_\ppm H_\mmp \rrb . \nonumber
\end{align}
The first bracket of the right hand side is the usual $T\bar T$ deformation. To understand how the second bracket changes the energy level, we consider the two-point correlation function.
\begin{align}
    \mathcal{C} = \ld H_\ppp(x)G_\mmm(\xp) \rd 
    & - \ld G_\ppp(x)H_\mmm(\xp) \rd \\
    & - \ld H_\ppm(x)G_\mmp(\xp) \rd + \ld G_\ppm(x)H_\mmp(\xp) \rd  . \nonumber
\end{align}
Up to contact terms that vanish at separated points, we can replace $G$ by using the conservation equation (\ref{equation:Conservation G H}):
\begin{align}
    \mathcal{C} = \ld H_\ppp(x)\partial^\prime_\mm H_{\mmp}(\xp) \rd 
    &- \ld \partial_\pp H_\ppm(x)  H_\mmm(\xp) \rd \\
    &- \ld H_\ppm(x)\partial^\prime_\pp H_\mmm(\xp) \rd + \ld \partial_\mm H_\ppp(x)H_\mmp(\xp) \rd . \nonumber
\end{align}
Here $\partial^\prime$ means the derivative with respect to the coordinate $\xp$. Now we can use translational invariance to move the derivative from $\xp$ to $x$. Then the first term cancels the fourth term and the third term cancels the second one because both $H$ and $G$ are fermionic, hence $\mathcal{C}$ vanishes at separated points. This implies the extra term can have no effect on the energy level. The presence of the extra term is only to make the action supersymmetric. The theory remains solvable, like the usual $T\bar T$ deformation, with the same relation between deformed and undeformed energy levels.

\section{Theories with \texorpdfstring{$(1,1)$}{Lg} Supersymmetry}\label{section:(1,1)}
% Example of deformed (1,1) free and interacting theory.

In this section, we consider the supercurrent-squared deformation of a theory involving a single $(1,1)$ superfield $\Phi$, both in the free case and with a superpotential.

\subsection{Free \texorpdfstring{$(1,1)$}{Lg} superfield}

First consider an undeformed superspace Lagrangian $\mathcal{A}^{(0)} = D_+ \Phi D_- \Phi$. We make the following ansatz for the deformed Lagrangian at finite $t$:
\begin{align}
    \mathcal{A}^{(t)} = F \left( \, t \partial_{++} \Phi \partial_{--} \Phi , \, t \left( D_+ D_- \Phi \right)^2 \right) D_+ \Phi D_- \Phi .
\end{align}
Here $F$ may only depend on the two dimensionless combinations which we define by 
\begin{align}
   x = t\, \partial_{++} \Phi \partial_{--} \Phi, \qquad  y = t \left( D_+ D_- \Phi \right)^2.
\end{align} In order to reduce to the free theory as $t \to 0$, we must also impose the boundary condition $F(0,0) = 1$.

After computing the components of the supercurrent-squared deformation and simplifying, the flow equation (\ref{ttbar_general_flow}) yields
\begin{align}
\begin{split}
    \frac{\partial}{\partial t} F &= \left( \left( D_+ D_- \Phi \right)^2 - \partial_{++} \Phi \partial_{--} \Phi \right) F^2 \\
    &\quad - 2 F \left( \partial_{++} \Phi \partial_{--} \Phi \right) \left( \partial_{++} \Phi \partial_{--} \Phi + \left( D_+ D_- \Phi \right)^2 \right) \frac{\partial F}{\partial x} .
    \label{free_pde}
\end{split}
\end{align}
In terms of the dimensionless variables $x$ and $y$, equation (\ref{free_pde}) becomes
\begin{align}
    \frac{\partial F}{\partial x} x + \frac{\partial F}{\partial y} y = (y-x)F^2 - 2 F \frac{\partial F}{\partial x} x ( x + y ) .
    \label{free_pde_nondim}    
\end{align}
Supplemented with the boundary condition $F(0,0) = 1$, the partial differential equation (\ref{free_pde}) uniquely determines the deformed Lagrangian at finite $t$.

As a check, we would like to verify that the bosonic structure of the solution to (\ref{free_pde}) reduces to the known results for the $T \Tb$-deformed theory of a free boson. We will argue that, in fact, it suffices to set $y = 0$ in (\ref{free_pde}) and note that the result agrees with the flow equation obtained in the purely bosonic case \cite{cavagliaBarTDeformed2D2016}.

Indeed, let us write the components of the superfield $\Phi$ as $\Phi = \phi + i \theta^+ \psi_+ + i \theta^- \psi_- + \theta^+ \theta^- f$. To probe the bosonic structure, it suffices to set $\psi_{\pm} = 0$, perform the superspace integration, and then integrate out the auxiliary field $f$ using its equation of motion. Thus consider an arbitrary superspace integral of the form
\begin{align}
    \mathcal{L}^{(t)} = \int d^2 \theta F^{(t)} (x, y) D_+ \Phi D_- \Phi .
\end{align}
The lowest component of the superfield $y = t D_+ \Phi D_- \Phi$ is $-f$, and the higher components will not contribute to the bosonic part because they come multiplying $D_+ \Phi D_- \Phi$, which is already proportional to $\theta^+ \theta^-$ after setting the fermions to zero.

Thus the purely bosonic piece of the physical Lagrangian associated with a superspace Lagrangian $\mathcal{A}^{(t)} = F^{(t)} ( x, y) D_+ \Phi D_- \Phi$ is
\begin{align}
    \mathcal{L}^{(t)} = F^{(t)} \left( t \partial_{++} \phi \partial_{--} \phi , t f^2 \right) \left( f^2 + 4 \partial_{++} \phi \partial_{--} \phi \right) .
\end{align}
The equation of motion for the auxiliary field $f$ is
\begin{align}
    2 t f \frac{\partial F}{\partial y} \left( f^2 + 4 \partial_{++} \phi \partial_{--} \phi \right) + 2 f F = 0 ,
\end{align}
which admits the solution $f=0$. The Lagrangian for the bosonic field $\phi$ is then
\begin{align}
    \mathcal{L}^{(t)} = 4 F^{(t)} \left( t \partial_{++} \phi \partial_{--} \phi , 0 \right) \partial_{++} \phi \partial_{--} \phi .
\end{align}
Therefore, to determine the terms in the Lagrangian which involve only $\phi$, we may solve the simpler partial differential equation
\begin{align}\label{equation:free_pde_y=0}
    \frac{\partial F}{\partial x} x = - x F^2 - 2 F x^2 \frac{\partial F}{\partial x},
\end{align}
which holds upon setting $y=0$ in (\ref{free_pde_nondim}). But this is precisely the equation discussed in section \ref{TTb-review}, whose solution is equation (\ref{equation:Nambu-Goto}):
\begin{align}
    \mathcal{L}^{(t)} = \frac{\sqrt{1 + 4 t \partial_{++} \phi \partial_{--} \phi} - 1}{2 t} . 
\end{align}
We see that the supercurrent-squared deformation of the free superfield is indeed a generalization of the $T \Tb$ deformation of a free boson, in the sense that it yields the same modification to the purely bosonic terms in the action but also includes additional terms which affect only the fermions.

\subsection{Interacting \texorpdfstring{$(1,1)$}{Lg} Superfield}

Next, we consider the case with a superpotential: that is, we begin from the undeformed superspace Lagrangian
\begin{align}
    \mathcal{A}^{(0)} = D_+ \Phi D_- \Phi + h ( \Phi ) , 
\end{align}
where $h(\Phi)$ is an arbitrary function (it need not give rise to a theory with infinitely many integrals of motion). After performing the superspace integral, the physical Lagrangian is
\begin{align}
    \mathcal{L}^{(0)} = \int d^2 \theta \, \mathcal{A}^{(0)} = \partial_{++} \phi \partial_{--} \phi + \psi_+ \partial_{--} \psi_+ + \psi_- \partial_{++} \psi_- + f^2 + h'(\phi) f .
\end{align}
Integrating out the auxiliary field using its equation of motion $f = - \frac{1}{2} h'(\phi)$, we see that the physical potential $V$ is given by $V = - \frac{1}{4} h'(\phi)^2$.

We might expect that both the kinetic and potential terms are modified by a finite supercurrent-squared deformation, which would lead us to make the ansatz
\begin{align}
    \mathcal{A}^{(t)} = F(x, y) D_+ \Phi D_- \Phi + G(t, \Phi) ,
    \label{potential_ansatz}
\end{align}
where $G$ is a new function to be determined, and $x = t \partial_{++} \Phi \partial_{--} \Phi$, $y = t \left( D_+ D_- \Phi \right)^2$ as above. However, the deformation does not induce any change in the potential $h$, so in fact we may put $G = h$ for all $t$. To see this, we can write down the supercurrent-squared deformation associated with the ansatz (\ref{potential_ansatz}), which gives
\begin{align}
%\begin{split}
	& \frac{\partial}{\partial t} F ( x, y ) D_+ \Phi D_- \Phi + \frac{\partial}{\partial t} G (t,\Phi) = \label{interacting_deformation} \\
	&\frac{1}{t} \Bigg( \left( y - x \right) F^2 - 2 F x ( x + y ) \frac{\partial F}{\partial x} + \left( G' \right)^2 + 2 G' \sqrt{y} \left( x \frac{\partial F}{\partial x} - F \right) - 2 \sqrt{y} x G' \frac{\partial F}{\partial y} \Bigg) D_+ \Phi D_- \Phi . \non
%\end{split}
\end{align}
The details of the calculation leading to (\ref{interacting_deformation}) are discussed in Appendix \ref{11_appendix}. We see that deformation is proportional to $D_+ \Phi D_- \Phi$, so it does not source any change in the potential $h(\Phi)$; thus we may take $G(h, \Phi) = h(\Phi)$ in our ansatz. This leaves us with a single partial differential equation for $F$, namely
\begin{align}
	 x \frac{\partial F}{\partial x} + y \frac{\partial F}{\partial y} &= \left( y - x \right) F^2 - 2 F x ( x + y ) \frac{\partial F}{\partial x} + \left( h' \right)^2 + 2 h' \sqrt{y} \left( x \frac{\partial F}{\partial x} - F \right) - 2 \sqrt{y} x h' \frac{\partial F}{\partial y} .
    \label{potential_pde}
\end{align}	
In the second line, we have used the constraint that $F$ can depend only on the dimensionless combinations $x = t \partial_{++} \Phi \partial_{--} \Phi$ and $y = t \left( D_+ D_- \Phi \right)^2$.

As in the free case, we would like to study the purely bosonic terms in the physical Lagrangian resulting from (\ref{potential_pde}) and compare them to known results. Here the auxiliary will play a more important role since $f=0$ is no longer a solution. 

We can expand both the Lagrangian $\mathcal{L} = \int d^2 \theta \left( F(x,y) D_+ \Phi D_- \Phi + h ( \Phi ) \right)$ and the auxiliary field $f$ as power series in $t$:
\begin{align}
    \mathcal{L} = \sum_{j=0}^{\infty} t^j \mathcal{L}^{(j)}, \qquad f = \sum_{j = 0}^{\infty} t^j f^{(j)} , 
\end{align}
and then integrate out the auxiliary order-by-order in $t$. Doing so to order $t^3$, we arrive at 
\begin{align}
\begin{split}
    \mathcal{L}  &= -\frac{1}{4} h'(\phi )^2 + \frac{x}{t} +t \left(\frac{1}{16} h'(\phi )^4 - \left( \frac{x}{t} \right)^2\right) +t^2 \left(-\frac{1}{4} \left( \frac{x}{t} \right)^2 h'(\phi )^2-\frac{1}{64} h'(\phi )^6+2 \left( \frac{x}{t} \right)^3\right) \label{our_result_expanded} \\
    &\quad + t^3 \left(\left( \frac{x}{t} \right)^3 h'(\phi )^2+\frac{1}{256} h'(\phi )^8-5 \left( \frac{x}{t} \right)^4\right) + \mathcal{O} ( t^4 ), 
\end{split}
\end{align}
after setting the fermions to zero. Up to conventions, this matches the Taylor expansion of the known result \cite{bonelliBarDeformationsClosed2018} for the $T \Tb$ deformation of a boson with a generic potential $V$, which is given in our conventions as
\begin{align}
    \mathcal{L}^{(t)} = -\frac{1}{2t} \frac{1-2 t V}{1-t V} + \frac{1}{2t} \sqrt{\frac{ t \left( 4 V + \partial_{++} \phi \partial_{--} \phi \right) }{1-t V}+\frac{(1-2 t V)^2}{(1-t V)^2}} .
    \label{closed_form_pot}
\end{align}
Again the physical potential $V$ is related to $h$ via $V = - \frac{1}{4} h'(\phi)^2$. We have checked explicitly that the bosonic part of the series solution to the PDE (\ref{potential_pde}) matches the Taylor expansion of (\ref{closed_form_pot}) up to $\mathcal{O} ( t^7 )$.

\section{Theories with \texorpdfstring{$(0,1)$}{Lg} Supersymmetry}\label{section:(0,1)}
% Example of deformed (0,1) free and interacting theory.
%
% CF: I added an i in the definition of the (0,1) superfield for easier comparison with reduction of (1,1).
%
In this section we study the deformation of a theory with chiral $(0,1)$ supersymmetry; a $(0,1)$ scalar superfield $\Phi$ consists of a scalar and a real fermion, $\Phi = \phi + i \tp \psi_+$. The Lagrangian in superspace is a function of $D_+ \Phi$, $\partial_\pp\Phi$, $\partial_\mm\Phi$, as well as $\Phi$ itself.

\subsection{Free $(0,1)$ superfield}

The free theory is defined by the Lagrangian,
\begin{align}
\begin{split}
    \mathcal{L} &= \int d\tp\ D_+\Phi\partial_\mm\Phi, \label{equation:(0,1) free} \\
    &= \partial_{++} \phi \partial_{--} \phi + \psi_+ \partial_{--} \psi_+ .
\end{split}
\end{align}
Following the approach of section \ref{section:supercurrent-squared}, we first look for conservation laws for a given superspace Lagrangian $\mathcal{A}$. They take the form, 
\begin{align}
\begin{split}
	& \partial_\mm \mathcal{S}_\ppp + D_+ \T_\ppmm = 0, \\
	& \partial_\mm \mathcal{S}_\mmp + D_+ \T_\mmmm = 0,
\end{split}
\end{align}
where $\mathcal{S}_{\pm\pm+}$ and $\T_{\ww\mm}$ are superfields given by:
\begin{align}\label{equation:(0,1) supercurrent}
\begin{split}
    & \mathcal{S}_\ppp = \frac{\delta\mathcal{A}}{\delta\partial_\mm \Phi}\partial_\pp \Phi, \\
    & \mathcal{S}_\mmp = \frac{\delta\mathcal{A}}{\delta\partial_\mm \Phi}\partial_\mm \Phi - \mathcal{A}, \\
    & \T_\ppmm = \frac{\delta\mathcal{A}}{\delta D_+ \Phi} \partial_\pp \Phi + D_+ \left( \frac{\delta\mathcal{A}}{\delta\partial_\pp \Phi} \partial_\pp \Phi \right) -D_+ \mathcal{A}, \\
    & \T_\mmmm = \frac{\delta\mathcal{A}}{\delta D_+ \Phi} \partial_\mm \Phi + D_+ \left( \frac{\delta\mathcal{A}}{\delta\partial_\pp \Phi} \partial_\mm  \Phi \right).
\end{split}
\end{align}
We define the supercurrent-squared deformation as follows: 
\begin{align}\label{equation:(0,1) flow equation}
    \frac{\partial}{\partial t} \mathcal{A}^{(t)} = \mathcal{S}_\ppp \T_\mmmm - \mathcal{S}_\mmp\T_\ppmm .
\end{align}

To understand what the deformation (\ref{equation:(0,1) flow equation}) does to a $(0,1)$ theory, consider an undeformed Lagrangian in superspace
\begin{align}
    \mathcal{A}^{(0)} = g(\Phi) D_+\Phi\partial_\mm\Phi, 
\end{align}
where $g(\Phi)$ is an arbitrary differentiable function of the superfield. A free theory corresponds to a constant $g(\Phi)$. To find the deformed theory $\mathcal{A}^{(t)}$, we first make a general ansatz for the deformed  Lagrangian 
\begin{align}
    \mathcal{A}^{(t)} = f(t \partial_\pp\Phi\partial_\mm\Phi)D_+\Phi\partial_\mm\Phi,
\end{align} 
where $f(x)$ is some differentiable function. Using the expression for the supercurrents given in (\ref{equation:(0,1) supercurrent}) and imposing the initial condition $f(x\rightarrow 0) = g(\Phi)$, we find the function $f(x)$ satisfies the same differential equation found in (\ref{equation:free_pde_y=0}). Its solution is given by
\begin{align}
    f(x) = \frac{\sqrt{1+4x g(\Phi)}-1}{2x}.
\end{align}

\subsection{Reduction of $(1,1)$ to $(0,1)$}

Any theory with $(1,1)$ global supersymmetry can also be viewed as a theory with $(0,1)$ global supersymmetry. Up to possible field redefinitions, we should therefore be able to relate the $(1,1)$ theory deformed by the supercurrent-squared deformation defined in (\ref{final_ttbar_general}) to the $(0,1)$  of (\ref{equation:(0,1) flow equation}), which we would have used if we had simply restricted to $(0,1)$ supersymmetry. 

%Therefore, one must be able to deform a $(1,1)$ theory by the supercurrent-squared deformation defined in (\ref{final_ttbar_general}) and view it as a $(0,1)$ theory, which should recover the flow expressed by equation (\ref{equation:(0,1) flow equation}) .

To be more precise, consider a $(1,1)$ theory whose physical Lagrangian $\mathcal{L}$ is given by the integral of a superspace Lagrangian $\mathcal{A}^{(1,1)}$ over $(1,1)$ superspace. We can also view this as a $(0,1)$ theory, 
\begin{align}
    \mathcal{L} = \int d^2 \theta \, \mathcal{A}^{(1,1)} = \int d \theta^+ \, \mathcal{A}^{(0,1)} . 
    \label{11-to-01}
\end{align}
The flow equation defining the supercurrent-squared deformation of $\mathcal{A}^{(1,1)}$ is $\partial_{t} \mathcal{A}^{(1,1)} = \mathcal{T}_{+++} \mathcal{T}_{---} - \mathcal{T}_{--+} \mathcal{T}_{++-}$. By performing the integral over $\theta^-$, this induces a flow for $\mathcal{A}^{(0,1)}$ due to (\ref{11-to-01}), namely
\begin{align}
    \frac{\partial}{\partial t} \mathcal{A}^{(0,1)} &= \int d \theta^+ \, \left( \mathcal{T}_{+++} \mathcal{T}_{---} - \mathcal{T}_{--+} \mathcal{T}_{++-} \right) .
    \label{reduced_flow}
\end{align}
For instance, let us consider the deformation of the free theory $\mathcal{A}^{(1,1)} = D_+ \Phi^{(1,1)} D_- \Phi^{(1,1)}$. This can be written as an integral over $(0,1)$ superspace as
\begin{align}
\begin{split}
    \int d^2 \theta \, D_+ \Phi^{(1,1)} D_- \Phi^{(1,1)} &= \int d \theta^+ \Big( - i \psi_+ \partial_{--} \phi - i \psi_- f - \theta^+ \big( f^2 + \partial_{++} \phi \partial_{--} \phi \label{11-reduced} \\
    &\hspace{70pt}  + \psi_+ \partial_{--} \psi_+ + \psi_- \partial_{++} \psi_- \big) \Big) \\
    &= - \int d \theta^+ \, \left(D_+ \Phi^{(0,1)} \partial_{--} \Phi^{(0,1)} + \Psi_- D_+ \Psi_- \right)  .
\end{split}
\end{align}
Here we have written the integrand on the right side of (\ref{11-reduced}) as a superspace Lagrangian $\mathcal{A}^{(0,1)} \left( \Phi^{(0,1)}, \Psi_- \right)$ for a superfield $\Phi^{(0,1)} = \phi + i \theta^+ \psi_+$ of the form discussed above, along with an extra Fermi superfield $\Psi_- = i \psi_- + \theta^+ f$:
\begin{align}
    \mathcal{A}^{(0,1)} \left( \Phi^{(0,1)}, \Psi_- \right) = D_+ \Phi^{(0,1)} \partial_{--} \Phi^{(0,1)} + \Psi_- D_+ \Psi_- .
\end{align}

For comparison, we compute the supercurrent-squared deformation to leading order in $t$; that is, we compute the tangent vector $\frac{\partial \mathcal{A}^{(1,1)}}{\partial t} \vert_{t=0}$ to the free theory along the flow and compare it to that of the free $(0,1)$ theory with an extra fermion.

The components of the supercurrent superfield associated with the free theory, after integrating out the auxiliary using $f=0$, are given in equation (\ref{sc-square-no-f}). Using these and performing the integral over $\theta^-$, the reduced flow equation (\ref{reduced_flow}) at $t=0$ becomes
\begin{align}
\begin{split}
    \frac{\partial}{\partial t} \mathcal{A}^{(0,1)} \vert_{t=0} &=  i \left( \psi_+ \partial_{++} \phi T_{----} + \psi_+ \psi_- \partial_{++} \psi_- \partial_{--} \phi \right) + \theta^+ \Big( T_{++++} T_{----} \label{bos-T-free} \\
    &\quad + 2 \partial_{++} \phi \partial_{--} \phi \left( \psi_+ \partial_{--} \psi_+  +  \psi_- \partial_{++} \psi_- \right) + \psi_- \partial_{++} \psi_- \psi_+ \partial_{--} \psi_+ \Big), 
\end{split}
\end{align}
where we have used $T_{\pm \pm \pm \pm} = \left( \partial_{\pm \pm} \phi \right)^2 + \psi_\pm \partial_{\pm \pm} \psi_\pm$.

We know that the solution to (\ref{bos-T-free}) must represent a solvable deformation of the original $(0,1)$ theory because it descends from a solvable deformation in the parent $(1,1)$ theory. On the other hand, one can construct the flow equation (\ref{equation:(0,1) flow equation}) directly in the $(0,1)$ theory. This must also yield a solvable deformation since it is built out of currents which satisfy a superspace conservation equation of the form used in the proof of section (\ref{section:solvable}). One might suspect that these two deformations should be the same, up to field redefinitions which do not affect the spectrum.

To check this, let us compare the leading-order deformations for these two cases in components. After including the contributions $\partial_{\pm \pm} \Psi_- \frac{\delta \mathcal{A}}{\delta D_{\pm} \Psi_-} $ to $\mathcal{T}_{\pm \pm - -}$ due to the fermion $\Psi_-$, the currents (\ref{equation:(0,1) supercurrent}) for this theory are
\begin{align}
\begin{split}
    \mathcal{S}_\ppp &= D_+ \Phi \partial_{++} \Phi \\
    &= i \psi_+ \partial_{++} \phi + \theta^+ \left( \psi_+ \partial_{++} \psi_+ + \left( \partial_{++} \phi \right)^2 \right) , \\
    \mathcal{S}_\mmp &= - \Psi_- D_+ \Psi_- \\
    &= - i \psi_- f - \theta^+ \left( f^2 + \psi_- \partial_{++} \psi_- \right) , \\
    \T_\ppmm &= \partial_{--} \Phi \partial_{++} \Phi + \Psi_- \partial_{++} \Psi_- - D_+ \left( D_+ \Phi \partial_{--} \Phi + \Psi_- D_+ \Psi_- \right)  \\
    &=  - \psi_+ \partial_{--} \psi_+ - f^2 + i \theta^+ \left( \partial_{++} \phi \partial_{--} \psi_+ - \psi_+ \partial_{++} \partial_{--} \phi \right), \\
    \T_\mmmm &= \left( \partial_{--} \Phi \right)^2 - \Psi_- \partial_{--} \Psi_- \\
    &= \left( \partial_{--} \phi \right)^2 + \psi_- \partial_{--} \psi_- + i \theta^+ \left( - f \partial_{--} \psi_- + \psi_- \partial_{--} f + 2 \partial_{--} \phi \partial_{--} \psi_+ \right) .
\end{split}
\end{align}
The bilinears appearing in the $(0,1)$ deformation are
\begin{align}
\begin{split}
    \mathcal{S}_{+++} \mathcal{T}_{----} &= i \psi_+ \partial_{++} \phi T_{----} + \theta^+ \big( T_{++++} T_{----} \\
    &\hspace{70pt} + \psi_+ \partial_{++} \phi \left( \psi_- \partial_{--} f - f \partial_{--} \psi_- + 2 \partial_{--} \phi \partial_{--} \psi_+ \right) \big) , \\
    \mathcal{S}_{--+} \mathcal{T}_{++--} &= i \psi_- f \left( \psi_+ \partial_{--} \psi_+ + f^2 \right) + \theta^+ \big( \left( f^2 + \psi_- \partial_{++} \psi_- \right) \left( f^2 + \psi_+ \partial_{--} \psi_+ \right) \\
    &\hspace{70pt} + \psi_- f \left( \psi_+ \partial_{++} \partial_{--} \phi - \partial_{++} \phi \partial_{--} \psi_+ \right) \big),
\end{split}
\end{align}
and thus the $(0,1)$ flow equation at $t=0$ is given by
\begin{align}
%\begin{split}
    \frac{\partial}{\partial t} \mathcal{A}^{(0,1)} \vert_{t=0} &= \mathcal{S}_{+++} \mathcal{T}_{----} - \mathcal{S}_{--+} \mathcal{T}_{++--} \label{01_free_flow} \cr
    &= i \psi_+ \partial_{++} \phi T_{----} - i \psi_- f \left( \psi_+ \partial_{--} \psi_+ + f^2 \right) \\
    &+ \theta^+ \Big( T_{++++} T_{----} + \psi_+ \partial_{++} \phi \left( \psi_- \partial_{--} f - f \partial_{--} \psi_- + 2 \partial_{--} \phi \partial_{--} \psi_+ \right) \cr
    &- \left( f^2 + \psi_- \partial_{++} \psi_- \right) \left( f^2 + \psi_+ \partial_{--} \psi_+ \right)  - \psi_- f \left( \psi_+ \partial_{++} \partial_{--} \phi - \partial_{++} \phi \partial_{--} \psi_+ \right) \Big). \non
%\end{split}
\end{align}
The deformations (\ref{bos-T-free}) and (\ref{01_free_flow}) agree up to terms proportional to the equations of motion $f=0$ and $\partial_{++} \psi_- = 0$. At this order in $t$, such terms can be removed by making a field redefinition involving $f$ and $\psi_-$. If this can be repeated order-by-order in $t$, as we suspect should be the case, then the two flows are genuinely equivalent and give rise to deformed theories with the same energies.

\section*{Acknowledgements}

It is our pleasure to thank D.~Kutasov for helpful discussions and suggestions, and T.~Maxfield for both helpful discussions and for sharing his notes on two-dimensional supercurrent multiplets. C.~F. acknowledges support from the University of Chicago's divisional MS-PSD program. C.~C., C.~F. and S.~S. are supported in part by NSF Grant No. PHY1720480.

\appendix
\section{Details of the $(1,1)$ PDE Calculation}\label{11_appendix}

In this Appendix, we show some steps of the calculation which leads to the partial differential equation (\ref{potential_pde}) defining the supercurrent-squared deformation of a free theory with a potential. By setting $h=0$, this calculation also reproduces the PDE (\ref{free_pde}) which describes deformations of the free theory.

We would like to consider what happens when we deform the superspace Lagrangian $\mathcal{A}^{(0)} = D_+ \Phi D_- \Phi + h( \Phi )$, according to the flow equation (\ref{ttbar_general_flow}),
\begin{align*}
	\frac{\partial}{\partial t} \mathcal{A}^{(t)} = \mathcal{T}_{+++}^{(t)} \mathcal{T}_{---}^{(t)} - \mathcal{T}_{--+}^{(t)} \mathcal{T}_{++-}^{(t)}  .
\end{align*}
It will help to introduce some shorthand: we define $A = D_+ \Phi D_- \Phi$ so that $\mathcal{A}^{(0)} = A$, and let $x = t \partial_{++} \Phi \partial_{--} \Phi$ and $y  = t \left( D_+ D_- \Phi \right)^2$ as before. Also define the dimensionful combinations 
\begin{align} 
X = \partial_{++} \Phi \partial_{--} \Phi = \frac{x}{t}, \qquad Y = \left( D_+ D_- \Phi \right)^2 = \frac{y}{t}.
\end{align}
Our ansatz for the superspace Lagrangian at finite $t$ will be $\mathcal{A}^{(t)} = F ( x , y ) A + h(\Phi)$.

With this ansatz, some of the terms in (\ref{final_ttbar_general}) will not contribute to the right side of (\ref{ttbar_general_flow}). For instance, the terms $\frac{\delta \mathcal{A}}{\delta D_+ D_- \Phi} D_{\pm} \partial_{\pm} \Phi$ will be proportional to $D_+ \Phi D_- \Phi = A$. However, every term in the superspace supercurrent is proportional to $D_+ \Phi$, $D_- \Phi$, or $D_+ \Phi D_- \Phi$. Therefore, when we construct a bilinear in $\mathcal{T}$, any term containing $D_+ \Phi D_- \Phi$ will not contribute because it can only appear multiplying another term which contains at least one of $D_{\pm} \Phi$, which vanishes because $\left( D_{\pm} \Phi \right)^2 = 0$.

For our special ansatz, we will re-write the components of $\mathcal{T}$ keeping only terms which contribute to bilinears,
\begin{align}
\begin{split}
	\mathcal{T}_{++-} &\sim \partial_{++} \Phi \frac{\delta \mathcal{A}}{\delta D_+ \Phi} + \partial_{++} \Phi D_+ \left(  \frac{\delta \mathcal{A}}{\delta \partial_{++} \Phi} \right) - \frac{1}{2} \partial_{++} \Phi D_- \left( \frac{\delta \mathcal{A}}{\delta D_+ D_- \Phi} \right) - D_+ \mathcal{A}  , \label{final_ttbar_equiv} \\	
    \mathcal{T}_{+++} &\sim \partial_{++} \Phi \frac{\delta \mathcal{A}}{\delta D_- \Phi} + \partial_{++} \Phi D_- \left(  \frac{\delta \mathcal{A}}{\delta \partial_{--} \Phi} \right) + \frac{1}{2} \partial_{++} \Phi D_+ \left( \frac{\delta \mathcal{A}}{\delta D_+ D_- \Phi} \right) ,  \\
    \mathcal{T}_{---} &\sim \partial_{--} \Phi \frac{\delta \mathcal{A}}{\delta D_+ \Phi} +  \partial_{--} \Phi D_+ \left( \frac{\delta \mathcal{A}}{\delta \partial_{++} \Phi} \right) - \frac{1}{2} \partial_{--} \Phi D_- \left( \frac{\delta \mathcal{A}}{\delta D_+ D_- \Phi} \right)  ,  \\
    \mathcal{T}_{--+} &\sim \partial_{--} \Phi \frac{\delta \mathcal{A}}{\delta D_- \Phi} + \partial_{--} \Phi  D_- \left( \frac{\delta \mathcal{A}}{\delta \partial_{--} \Phi} \right) + \frac{1}{2} \partial_{--} \Phi D_+ \left( \frac{\delta \mathcal{A}}{\delta D_+ D_- \Phi} \right) - D_- \mathcal{A} . 
\end{split}
\end{align}
The terms are
\begin{align*}
	D_+ \mathcal{A} &\sim F D_+ A + h'(\Phi) D_+ \Phi, \\
	D_- \mathcal{A} &\sim F D_- A + h'(\Phi) D_- \Phi, \\
    \partial_{++} \Phi \frac{\delta \mathcal{A}}{\delta D_+ \Phi} &\sim F \partial_{++} \Phi D_- \Phi, \\
    \partial_{++} \Phi \frac{\delta \mathcal{A}}{\delta D_- \Phi} &\sim - F \partial_{++} \Phi D_+ \Phi , \\
    \partial_{--} \Phi \frac{\delta \mathcal{A}}{\delta D_+ \Phi} &\sim F \partial_{--} \Phi D_- \Phi ,\\ 
    \partial_{--} \Phi \frac{\delta \mathcal{A}}{\delta D_- \Phi} &\sim - F \partial_{--} \Phi D_+ \Phi , \\
    \partial_{++} \Phi D_+ \left( \frac{\delta \mathcal{A}}{\delta \partial_{++} \Phi} \right) &\sim X \frac{\partial F}{\partial X} D_+ A ,\\
    \partial_{++} \Phi D_- \left( \frac{\delta \mathcal{A}}{\delta \partial_{--} \Phi} \right) &\sim \left( \partial_{++} \Phi \right)^2 \frac{\partial F}{\partial X} D_- A ,\\
    \partial_{--} \Phi D_+ \left( \frac{\delta \mathcal{A}}{\delta \partial_{++} \Phi} \right) &\sim \left( \partial_{--} \Phi \right)^2 \frac{\partial F}{\partial X} D_+ A ,\\
    \partial_{--} \Phi D_- \left( \frac{\delta \mathcal{A}}{\delta \partial_{--} \Phi} \right) &\sim X \frac{\partial F}{\partial X} D_- A , \\
	\frac{1}{2} \partial_{++} \Phi D_+ \left( \frac{\delta \mathcal{A}}{\delta D_+ D_- \Phi} \right) &\sim \sqrt{Y} \partial_{++} \Phi \frac{\partial F}{\partial Y} \cdot D_+ A, \\
	\frac{1}{2} \partial_{--} \Phi D_+ \left( \frac{\delta \mathcal{A}}{\delta D_+ D_- \Phi} \right) &\sim \sqrt{Y} \partial_{--} \Phi \frac{\partial F}{\partial Y} \cdot D_+ A , \\
    - \frac{1}{2} \partial_{++} \Phi D_- \left( \frac{\delta \mathcal{A}}{\delta D_+ D_- \Phi} \right) &\sim - \sqrt{Y} \partial_{++} \Phi \frac{\partial F}{\partial Y} \cdot D_- A ,\\
    - \frac{1}{2} \partial_{--} \Phi D_- \left( \frac{\delta \mathcal{A}}{\delta D_+ D_- \Phi} \right) &\sim - \sqrt{Y} \partial_{--} \Phi \frac{\partial F}{\partial Y} \cdot D_- A , 
\end{align*}
where $\sim$ means ``equal modulo terms which are proportional to $D_+ \Phi D_- \Phi$,'' since any products involving these terms will contain two nilpotent factors and thus vanish.

The first piece of supercurrent-squared is
\begin{align}
\begin{split}
	\mathcal{T}_{++|+} \mathcal{T}_{--|-} &= \left( - F \partial_{++} \Phi D_+ \Phi + \left( \partial_{++} \Phi \right)^2 \frac{\partial F}{\partial X} D_- A + \sqrt{Y} \partial_{++} \Phi \frac{\partial F}{\partial Y} \cdot D_+ A \right) \\
    &\times \left( F \partial_{--} \Phi D_- \Phi +  \left( \partial_{--} \Phi \right)^2 \frac{\partial F}{\partial X} D_+ A - \sqrt{Y} \partial_{--} \Phi \frac{\partial F}{\partial Y} D_- A \right), \\
    &= - F^2 X A - F X \frac{\partial F}{\partial X} \partial_{--} \Phi D_+ \Phi D_+ A + F X \frac{\partial F}{\partial X} \partial_{++} \Phi D_- A D_- \Phi \\
    &+ X^2 \left( \frac{\partial F}{\partial X} \right)^2 D_- A D_+ A + F \frac{\partial F}{\partial Y} \sqrt{Y} X D_+ A D_- \Phi \\
    &+ F X \sqrt{Y} \frac{\partial F}{\partial Y} D_+ \Phi D_- A - Y X \left( \frac{\partial F}{\partial Y} \right)^2 D_+ A D_- A .
\end{split}
\end{align}
The second piece is
\begin{align}
\begin{split}
	\mathcal{T}_{++|-} \mathcal{T}_{--|+} &= \left( F \partial_{++} \Phi D_- \Phi + \left( X \frac{\partial F}{\partial X} - F \right) D_+ A - G' D_+ \Phi - \sqrt{Y} \partial_{++} \Phi \frac{\partial F}{\partial Y} D_- A \right) \\
    &\times \left(  - F \partial_{--} \Phi D_+ \Phi + \left( X \frac{\partial F}{\partial X} - F \right) D_- A - G' D_- \Phi + \sqrt{Y} \partial_{--} \Phi \frac{\partial F}{\partial Y} \cdot D_+ A \right), \\
    &= F^2 X A + F \left( X \frac{\partial F}{\partial X} - F \right) \partial_{++} \Phi D_- \Phi D_- A + F X \sqrt{Y} \frac{\partial F}{\partial Y} D_- \Phi D_+ A \\
    &+ F X \sqrt{Y} \frac{\partial F}{\partial Y} D_- A D_+ \Phi - F \left( X \frac{\partial F}{\partial X} - F \right) \partial_{--} \Phi D_+ A D_+ \Phi \\
    &+ \left( X \frac{\partial F}{\partial X} - F \right)^2 D_+ A D_- A - Y X \left( \frac{\partial F}{\partial Y} \right)^2 D_- A D_+ A \\
    &- G' \left( X \frac{\partial F}{\partial X} - F \right) D_+ \Phi D_- A + \left( G' \right)^2 D_+ \Phi D_- \Phi - G' \sqrt{Y} \partial_{--} \Phi \frac{\partial F}{\partial Y} D_+ \Phi D_+ A \\
    &- G' \left( X \frac{\partial F}{\partial X} - F \right) D_+ A D_- \Phi + G' \sqrt{Y} \partial_{++} \Phi \frac{\partial F}{\partial Y} D_- A D_- \Phi . 
    \label{potential_ttbar}
\end{split}
\end{align}
Using the definitions $A = D_+ \Phi D_- \Phi$, $X = \partial_{++} \Phi \partial_{--} \Phi$, and $\sqrt{Y} = D_+ D_- \Phi$, we see that the products appearing in the above bilinears can be simplified as follows:
\begin{align}
\begin{split}
	D_+ \Phi D_+ A &= D_+ \Phi D_+ \left( D_+ \Phi D_- \Phi \right) = D_+ \Phi D_+ D_+ \Phi D_- \Phi = A \partial_{++} \Phi , \\
    D_+ \Phi D_- A &= D_+ \Phi D_- \left( D_+ \Phi D_- \Phi \right) = D_+ \Phi D_- D_+ \Phi D_- \Phi = - A \sqrt{Y} , \\
    D_- \Phi D_+ A &= D_- \Phi D_+ \left( D_+ \Phi D_- \Phi \right) = - D_- \Phi D_+ \Phi D_+ D_- \Phi = A \sqrt{Y} , \\
    D_- \Phi D_- A &= D_- \Phi D_- \left( D_+ \Phi D_- \Phi \right) = - D_- \Phi D_+ \Phi D_- D_- \Phi = A \partial_{--} \Phi , \\
    D_+ A D_- A &= \left( \partial_{++} \Phi D_- \Phi - D_+ \Phi \sqrt{Y} \right) \left( - \sqrt{Y} D_- \Phi - \partial_{--} \Phi D_+ \Phi \right) = (X + Y) A .
\end{split}
\end{align}
So after simplifying,
\begin{align}
\begin{split}
	\mathcal{T}_{++|+} \mathcal{T}_{--|-} &= - F^2 X A - 2 F X^2 \frac{\partial F}{\partial X} A - X^2 \left( \frac{\partial F}{\partial X} \right)^2 A (X + Y) - 2 F \frac{\partial F}{\partial Y} Y X A , \label{sc-square-terms-potential} \\
	\mathcal{T}_{++|-} \mathcal{T}_{--|+} &= F^2 X A + 2 F X \left( X \frac{\partial F}{\partial X} - F \right) A + 2 F X Y \frac{\partial F}{\partial Y} A + \left( X \frac{\partial F}{\partial X} - F \right)^2 (X + Y) A \\
	&\quad + \left( \left( h' \right)^2 + 2 h' \sqrt{Y} \left( X \frac{\partial F}{\partial X} - F \right) - 2 \sqrt{Y} X h' \frac{\partial F}{\partial Y} \right) A .
\end{split}
\end{align}
In particular, we see that every term appearing in (\ref{sc-square-terms-potential}) is proportional to $A = D_+ \Phi D_- \Phi$. This means that the deformation only generates a change in the first term of our ansatz $\mathcal{A}^{(t)} = F D_+ \Phi D_- \Phi + h ( \Phi )$, but it does not source any change in the potential. This justifies our choice of ansatz which leaves the potential as $h(\Phi)$ rather than allowing a more general function $G(t, \Phi)$ with $G(0, \Phi) = h(\Phi)$.

Adding the contributions gives,
\begin{align}
\begin{split}
	\mathcal{T}_{++|+} \mathcal{T}_{--|-} +  \mathcal{T}_{++|-} \mathcal{T}_{--|+} &= \Big[ \left( Y - X \right) F^2 - 2 F X ( X + Y ) \frac{\partial F}{\partial X} + 2 h' \sqrt{Y} \left( X \frac{\partial F}{\partial X} - F \right) \\
	&\quad - 2 \sqrt{Y} X h' \frac{\partial F}{\partial Y} + \left( h' \right)^2 \Big] A .
\end{split}
\end{align}
Setting this deformation equal to $\frac{\partial}{\partial t} \mathcal{A}^{(t)}$, and multiplying both sides by $t$ to convert dimensionlful variables $X$ and $Y$ into their dimensionless counterparts $x$ and $y$, gives our final result (\ref{potential_pde}), 
\begin{align}
	x \frac{\partial}{\partial x} F + y \frac{\partial}{\partial y} F &= \left( y - x \right) F^2 - 2 F x ( x + y ) \frac{\partial F}{\partial x} + \left( h' \right)^2 + 2 h' \sqrt{y} \left( x \frac{\partial F}{\partial x} - F \right) \cr & - 2 \sqrt{y} x h' \frac{\partial F}{\partial y} . 
	\label{pde_in_appendix}
\end{align}
We were unable to find a closed-form solution to (\ref{potential_pde}) in the general case. However, we can find the solution in a few special cases. If $y=0$, (\ref{pde_in_appendix}) reduces to
%We were unable to write down the closed-form solution to (\ref{potential_pde}). However, we can find the solution in a few special cases. If $y=0$, (\ref{pde_in_appendix}) reduces to
%
\begin{align}
    x F'(x) = - x \left( F(x)^2 + 2 F(x) F'(x) x \right) ,
\end{align}
which is solved by the Dirac-type ansatz $F(x) = \frac{\sqrt{1+4x}-1}{2x}$. If $x=0$, equation (\ref{pde_in_appendix}) is solved by $F(y) = \frac{1}{1-y}$. If $y = - x$, the second term on the right side of (\ref{pde_in_appendix}) drops out and the solution is $F(x) = \frac{1}{1+2x}$.

\section{Details of the $(0,1)$ PDE Calculation}\label{01-appendix}

We would like to write down a partial differential equation, similar to (\ref{free_pde}) in the $(1,1)$ case, which determines the Lagrangian deformed by the $(0,1)$ supercurrent-squared at finite $t$.

Define the three combinations of fields 
\begin{align} 
x = t \partial_{++} \Phi \partial_{--} \Phi, \qquad  y = t \left( D_+ \Psi_- \right)^2, \qquad z = t D_+ \Phi D_+ \partial_{--} \Phi,
\end{align}
and their dimensionful counterparts $X = \frac{x}{t}$, $Y = \frac{y}{t}$, $Z = \frac{z}{t}$. Our ansatz for the Lagrangian at finite $t$ is
\begin{align}
    \mathcal{A}^{(t)} = F ( x, y, z) \left( D_+ \Phi \partial_{--} \Phi + \Psi_- D_+ \Psi_- \right) + F_{2, -} ( x, y, z ) \left( \Psi_- D_+ \Phi \right) . 
\end{align}
Since the function $F_{2,-}$ is fermionic, it actually contains several different functions since we may combine the fields $\Phi, \Psi_-$ and derivatives in a few independent ways to obtain a fermionic function. We will expand $F_{2,-}$ as follows:
\begin{align}
    F_{2,-} = G ( x, y, z ) D_+ \Psi_- D_+ \partial_{--} \Phi + H ( x, y, z ) \partial_{++} \Phi \partial_{--} \Psi_- + J (x, y, z) \partial_{--} \Phi \partial_{++} \Psi_- .
\end{align}
%
%Renaming $F_{1} \to F$, altogether our ansatz for the deformed Lagrangian is
Altogether our ansatz for the deformed Lagrangian is, 
\begin{align}
    \mathcal{A}^{(t)} &= F \left( D_+ \Phi \partial_{--} \Phi + \Psi_- D_+ \Psi_- \right) \nonumber \\
    &\quad + \left( G D_+ \Psi_- D_+ \partial_{--} \Phi + H \partial_{++} \Phi \partial_{--} \Psi_- + J \partial_{--} \Phi \partial_{++} \Psi_- \right) \left( \Psi_- D_+ \Phi \right) . 
    \label{01-reduced-ansatz}
\end{align}
This is a $(0,1)$ superspace Lagrangian with the functional dependence
\begin{align}
    \mathcal{A} = \mathcal{A} \left( \Phi, \Psi_-, D_+ \Phi, D_+ \Psi_-, \partial_{\pm \pm} \Phi, \partial_{\pm \pm} \Psi_-, D_+ \partial_{--} \Phi \right) . 
\end{align}
Following the procedure of section (\ref{section:supercurrent-squared}), we can consider a transformation $x^{\pm \pm} \to x^{\pm \pm} + a^{\pm \pm}$ and extract the components of conserved currents. In this case, they are
\begin{align}
\begin{split}
    \mathcal{T}_{++++} &= \partial_{++} \Phi \frac{\delta \mathcal{A}}{\delta \partial_{--} \Phi} + \partial_{++} \Psi_- \frac{\delta \mathcal{A}}{\delta \partial_{++} \Psi_-} - \partial_{++} \Phi D_+ \left( \frac{\delta \mathcal{A}}{\delta D_+ \partial_{--} \Phi} \right) , \\
    \mathcal{T}_{++--} &= \partial_{--} \Phi \frac{\delta \mathcal{A}}{\delta \partial_{--} \Phi} + \partial_{--} \Psi_- \frac{\delta \mathcal{A}}{\delta \partial_{--} \Psi_-} - \partial_{--} \Phi D_+ \left( \frac{ \delta \mathcal{A}}{\delta D_+ \partial_{--} \Phi} \right) - \mathcal{A} , \\
    \mathcal{S}_{++-} &= \partial_{++} \Phi \frac{\delta \mathcal{A}}{\delta D_+ \Phi} + D_+ \left( \partial_{++} \Phi \frac{\delta \mathcal{A}}{\delta \partial_{++} \Phi} - \mathcal{A} \right) + \partial_{++} \Psi_- \frac{\delta \mathcal{A}}{\delta D_+ \Psi_-}  \cr & + \left( \partial_{--} \partial_{++} \Phi \right) \frac{\delta \mathcal{A}}{\delta D_+ \partial_{--} \Phi} , \\
    \mathcal{S}_{---} &= \partial_{--} \Phi \frac{\delta \mathcal{A}}{\delta D_+ \Phi} + D_+ \left( \partial_{--} \Phi \frac{\delta \mathcal{A}}{\delta \partial_{++} \Phi} \right) + \partial_{--} \Psi_- \frac{\delta \mathcal{A}}{\delta D_+ \Psi_-} + \partial_{--}^2 \Phi \frac{\delta \mathcal{A}}{\delta D_+ \partial_{--} \Phi} .
\end{split}
\end{align}
We compute each of these contributions. As before, we will drop terms which are proportional to $\Psi_- D_+ \Phi$, since every term in $\mathcal{S}$ and $\mathcal{T}$ is proportional to either $\Psi_-$ or to $D_+ \Phi$, so any terms involving both of these nilpotent factors will not contribute to bilinears. We will also introduce the shorthand $A = D_+ \Phi \partial_{--} \Phi$ and $B = \Psi_- D_+ \Psi_-$.

Doing this, we see that:
\begin{align}
    \partial_{++} \Phi \frac{\delta \mathcal{A}}{\delta \partial_{--} \Phi} &\sim F  D_+ \Phi \partial_{++} \Phi + \frac{\partial F}{\partial x} \left( \partial_{++} \Phi \right)^2 \left( A + B \right) , \label{01-reduced-components} \cr
    \partial_{--} \Phi \frac{\delta \mathcal{A}}{\delta \partial_{--} \Phi} &\sim F A + x \frac{\partial F}{\partial x} \left( A + B \right) , \cr
    \partial_{++} \Phi \frac{\delta \mathcal{A}}{\delta D_+ \Phi} &\sim \partial_{++} \Phi \frac{\partial F}{\partial z} \left( D_+ \partial_{--} \Phi \right) \left( A + B \right) + F x \cr
    &\quad - \partial_{++} \Phi \left( G D_+ \Psi_- D_+ \partial_{--} \Phi + H \partial_{++} \Phi \partial_{--} \Psi_- + J \partial_{--} \Phi \partial_{++} \Psi_- \right) \Psi_- , \cr
    \partial_{--} \Phi \frac{\delta \mathcal{A}}{\delta D_+ \Phi} &\sim \partial_{--} \Phi \frac{\partial F}{\partial z} \left( D_+ \partial_{--} \Phi \right) + F \left( \partial_{--} \Phi \right)^2 \cr
    &\quad - \partial_{--} \Phi \left( G D_+ \Psi_- D_+ \partial_{--} \Phi + H \partial_{++} \Phi \partial_{--} \Psi_- + J \partial_{--} \Phi \partial_{++} \Psi_- \right) \Psi_-, \cr
    \partial_{++} \Psi_- \frac{\delta \mathcal{A}}{\delta \partial_{++} \Psi_-} & \sim 0 , \cr
    \partial_{--} \Psi_- \frac{\delta \mathcal{A}}{\delta \partial_{--} \Psi_-} &\sim 0 , \cr
    D_+ \left( \partial_{++} \Phi \frac{\delta \mathcal{A}}{\delta \partial_{++} \Phi} \right) &\sim D_+ \left( x \frac{\partial F}{\partial x} ( A + B ) \right) + \Big( x \frac{\partial G}{\partial x} D_+ \Psi_- D_+ \partial_{--} \Phi + x \frac{\partial H}{\partial x} \partial_{++} \Phi \partial_{--} \Psi_- \cr
    &\hspace{50pt}  + x \frac{\partial J}{\partial x} \partial_{--} \Phi \partial_{++} \Psi_- \Big) \cdot \Big( D_+ \Psi_- D_+ \Phi - \Psi_- \partial_{++} \Phi \Big),  \cr
    D_+ \left( \partial_{--} \Phi \frac{\delta \mathcal{A}}{\delta \partial_{++} \Phi} \right) &\sim D_+ \left( \left( \partial_{--} \Phi \right)^2 \frac{\partial F}{\partial x} ( A + B ) \right) + \Big( \frac{\partial G}{\partial x} \left( \partial_{--} \Phi \right)^2 D_+ \partial_{--} \Phi \cr
    &+ \frac{\partial H}{\partial x} \partial_{--} \Phi \partial_{--} \Psi_- + \frac{\partial J}{\partial x} \left( \partial_{--} \Phi \right)^3 \partial_{++} \Psi_- \Big) \Big( D_+ \Psi_- D_+ \Phi - \Psi_- \partial_{++} \Phi \Big), \cr
    \partial_{++} \Phi D_+ \left( \frac{\delta \mathcal{A}}{\delta D_+ \partial_{--} \Phi} \right) &\sim \partial_{++} \Phi \frac{\partial F}{\partial z} \left( \partial_{++} \Phi \Psi_- D_+ \Psi_- - D_+ \Phi \left( D_+ \Psi_- \right)^2 \right), \cr
    \partial_{--} \Phi D_+ \left( \frac{\delta \mathcal{A}}{\delta D_+ \partial_{--} \Phi} \right) &\sim \partial_{--} \Phi \frac{\partial F}{\partial z} \left( \partial_{++} \Phi \Psi_- D_+ \Psi_- - D_+ \Phi \left( D_+ \Psi_- \right)^2 \right), \cr
    \partial_{++} \Psi_- \frac{\delta \mathcal{A}}{\delta D_+ \Psi_-} &\sim \partial_{++} \Psi_- \left( \frac{\partial F}{\partial y} ( A + B ) + F \Psi_- \right) , \cr
    \partial_{--} \Psi_- \frac{\delta \mathcal{A}}{\delta D_+ \Psi_-} &\sim \partial_{--} \Psi_- \left( \frac{\partial F}{\partial y} ( A + B ) + F \Psi_- \right) , \cr
    \partial_{--} \partial_{++} \Phi \frac{\delta \mathcal{A}}{\delta D_+ \partial_{--} \Phi} &\sim \partial_{--} \partial_{++} \Phi  \left( \frac{\partial F}{\partial z} D_+ \Phi (A + B) \right) \sim 0 , \cr
    \partial_{--}^2 \Phi \frac{\delta \mathcal{A}}{\delta D_+ \partial_{--} \Phi} &\sim \partial_{--}^2 \Phi \left( \frac{\partial F}{\partial z} D_+ \Phi ( A + B ) \right) \sim 0 , \cr
    D_+ \mathcal{A} &\sim \left( \frac{\partial F}{\partial x} D_+ x + \frac{\partial F}{\partial y} D_+ y + \frac{\partial F}{\partial z} D_+ z \right) ( A + B ) + F \left( D_+ A + D_+ B \right) \cr
    &+ \left( G D_+ \Psi_- D_+ \partial_{--} \Phi + H \partial_{++} \Phi \partial_{--} \Psi_- + J \partial_{--} \Phi \partial_{++} \Psi_- \right) \cr
    &\quad \times \left( D_+ \Psi_- D_+ \Phi - \Psi_- \partial_{++} \Phi \right).
\end{align}
We will argue that the coupled differential equations for $F, G, H$, and $J$ resulting from (\ref{01-reduced-components}) are consistent. This will be the case if they do not source any additional combinations of fields that do not appear in the ansatz (\ref{01-reduced-ansatz}).

The only thing that could spoil consistency is a $D_+ x$ term, since
\begin{align}
    D_+ x = \left( D_+ \partial_{++} \Phi \right) \partial_{--} \Phi + \partial_{++} \Phi D_+ \partial_{--} \Phi .
\end{align}
We have already allowed for dependence on $D_+ \partial_{--} \Phi$ in our Lagrangian, but terms proportional to $D_+ \partial_{++} \Phi$ are forbidden. We will show that, in the $\mathcal{S} \cdot \mathcal{T}$ deformation resulting from (\ref{01-reduced-components}), all $D_+ x$ terms drop out.

Tracking only the $D_+ x$ terms in bilinears, the supercurrent components are
\begin{align}
\begin{split}
    \mathcal{S}_{++-} &\sim x D_+ \left( \frac{\partial F}{\partial x} \right) \left( A + B \right) + \ldots , \\
    \mathcal{T}_{++--} &\sim x \frac{\partial F}{\partial x} ( A + B ) - F B - \partial_{--} \Phi \frac{\partial F}{\partial z} D_+ \left( D_+ \Phi \Psi_- D_+ \Psi_- \right) , \\
    \mathcal{S}_{---} &\sim \left( \partial_{--} \phi \right)^2 D_+ \left( \frac{\partial F}{\partial x} \right) ( A + B ) + \ldots , \\
    \mathcal{T}_{++++} &\sim F D_+ \Phi \partial_{++} \Phi + \frac{\partial F}{\partial x} \left( \partial_{++} \Phi \right)^2 ( A + B ) - \partial_{++} \Phi \frac{\partial F}{\partial z} D_+ \left( D_+ \Phi \Psi_- D_+ \Psi_- \right) ,
\end{split}
\end{align}
where $\ldots$ indicates terms that are not proportional to $D_+ x$ or $D_+ \left( \frac{\partial F}{\partial x} \right)$.

The relevant contributions in the deformation are
\begin{align}
\begin{split}
  &  \mathcal{T}_{++++} \mathcal{S}_{---} - \mathcal{T}_{++--} \mathcal{S}_{++-} \sim \left( \partial_{--} \Phi \right)^2 D_+ \left( \frac{\partial F}{\partial x} \right) ( A + B ) \cdot \left( F D_+ \Phi \partial_{++} \Phi + \frac{\partial F}{\partial x} \left( \partial_{++} \Phi \right)^2 (A + B ) \right) \\
    &\quad - x D_+ \left( \frac{\partial F}{\partial x} \right) (A + B ) \cdot \left( x \frac{\partial F}{\partial x} ( A + B ) - F B \right) + \ldots,
    \\
    &= \left( \partial_{--} \Phi \right)^2 D_+ \left( \frac{\partial F}{\partial x} \right) ( F B D_+ \Phi \partial_{++} \Phi ) - x D_+ \left( \frac{\partial F}{\partial x} \right) ( - F A B ) + \ldots , 
\end{split}
\end{align}
where we have used the fermionic nature of $A$ and $B$ so $A^2 = B^2 = (A+B)^2 = 0$. However in the last line, we recognize that $\left( \partial_{--} \Phi \right)^2 D_+ \Phi \partial_{++} \Phi = x A$, since $x = \partial_{++} \Phi \partial_{--} \Phi$ and $A = D_+ \Phi \partial_{--} \Phi$, so
\begin{align}
\begin{split}
    \mathcal{T}_{++++} \mathcal{S}_{---} - \mathcal{T}_{++--} \mathcal{S}_{++-} &\sim x D_+ \left( \frac{\partial F}{\partial x} \right) F B A + x D_+ \left( \frac{\partial F}{\partial x} \right) F A B = 0 , 
\end{split}
\end{align}
and thus the problematic $D_+ \left( \frac{\partial F}{\partial x} \right)$ terms do not contribute.

\newpage
\bibliographystyle{utphys}
\bibliography{master}

\providecommand{\href}[2]{#2}\begingroup\raggedright\begin{thebibliography}{10}

\bibitem{zamolodchikovExpectationValueComposite2004}
A.~B. Zamolodchikov, ``Expectation Value of Composite Field $T \Tb$ in
  Two-Dimensional Quantum Field Theory,''
  \href{http://www.arXiv.org/abs/hep-th/0401146}{{\tt hep-th/0401146}}.

\bibitem{bonelliBarDeformationsClosed2018}
G.~Bonelli, N.~Doroud, and M.~Zhu, ``$T \Tb$-Deformations in Closed Form,''
  {\em Journal of High Energy Physics} {\bf 2018} (June, 2018)
  \href{http://www.arXiv.org/abs/1804.10967}{{\tt 1804.10967}}.

\bibitem{cavagliaBarTDeformed2D2016}
A.~Cavagli\`a, S.~Negro, I.~M. Sz\'ecs\'enyi, and R.~Tateo, ``$T \Tb$-Deformed
  {{2D Quantum Field Theories}},'' {\em Journal of High Energy Physics} {\bf
  2016} (Oct., 2016) \href{http://www.arXiv.org/abs/1608.05534}{{\tt
  1608.05534}}.

\bibitem{Giveon:2017nie}
A.~Giveon, N.~Itzhaki, and D.~Kutasov, ``{$ \mathrm{T}\overline{\mathrm{T}} $
  and LST},'' {\em JHEP} {\bf 07} (2017) 122,
\href{http://www.arXiv.org/abs/1701.05576}{{\tt 1701.05576}}.
%%CITATION = ARXIV:1701.05576;%%.

\bibitem{Aharony:2018bad}
O.~Aharony, S.~Datta, A.~Giveon, Y.~Jiang, and D.~Kutasov, ``{Modular
  invariance and uniqueness of $T\bar{T}$ deformed CFT},''
\href{http://www.arXiv.org/abs/1808.02492}{{\tt 1808.02492}}.
%%CITATION = ARXIV:1808.02492;%%.

\bibitem{Aharony:2018ics}
O.~Aharony, S.~Datta, A.~Giveon, Y.~Jiang, and D.~Kutasov, ``{Modular
  covariance and uniqueness of $J\bar{T}$ deformed CFTs},''
\href{http://www.arXiv.org/abs/1808.08978}{{\tt 1808.08978}}.
%%CITATION = ARXIV:1808.08978;%%.

\bibitem{Taylor:2018xcy}
M.~Taylor, ``{TT deformations in general dimensions},''
\href{http://www.arXiv.org/abs/1805.10287}{{\tt 1805.10287}}.
%%CITATION = ARXIV:1805.10287;%%.

\bibitem{hartmanHolographyFiniteCutoff2018}
T.~Hartman, J.~Kruthoff, E.~Shaghoulian, and A.~Tajdini, ``Holography at Finite
  Cutoff with a $T^2$ Deformation,'' {\em arXiv:1807.11401 [hep-th]} (July,
  2018) \href{http://www.arXiv.org/abs/1807.11401}{{\tt 1807.11401}}.

\bibitem{cardyOverlineDeformationQuantum2018}
J.~Cardy, ``The $T \Tb$ Deformation of Quantum Field Theory as Random
  Geometry,'' {\em arXiv:1801.06895 [hep-th]} (Jan., 2018)
  \href{http://www.arXiv.org/abs/1801.06895}{{\tt 1801.06895}}.

\bibitem{dubovskyBarTPartitionFunction2018}
S.~Dubovsky, V.~Gorbenko, and G.~{Hernandez-Chifflet}, ``$T \Tb$ {{Partition
  Function}} from {{Topological Gravity}},'' {\em arXiv:1805.07386 [hep-th]}
  (May, 2018) \href{http://www.arXiv.org/abs/1805.07386}{{\tt 1805.07386}}.

\bibitem{Dubovsky2017}
S.~Dubovsky, V.~Gorbenko, and M.~Mirbabayi, ``Asymptotic Fragility, near
  {{AdS}}$ _2$ Holography and $T\Tb$,'' {\em Journal of High Energy Physics}
  {\bf 2017} (Sept., 2017) 136.

\bibitem{Caselle:2013dra}
M.~Caselle, D.~Fioravanti, F.~Gliozzi, and R.~Tateo, ``{Quantisation of the
  effective string with TBA},'' {\em JHEP} {\bf 07} (2013) 071,
\href{http://www.arXiv.org/abs/1305.1278}{{\tt 1305.1278}}.
%%CITATION = ARXIV:1305.1278;%%.

\bibitem{Baggio:2018gct}
M.~Baggio and A.~Sfondrini, ``{Strings on NS-NS Backgrounds as Integrable
  Deformations},'' {\em Phys. Rev.} {\bf D98} (2018), no.~2, 021902,
\href{http://www.arXiv.org/abs/1804.01998}{{\tt 1804.01998}}.
%%CITATION = ARXIV:1804.01998;%%.

\bibitem{Dei:2018mfl}
A.~Dei and A.~Sfondrini, ``{Integrable spin chain for stringy
  Wess-Zumino-Witten models},'' {\em JHEP} {\bf 07} (2018) 109,
\href{http://www.arXiv.org/abs/1806.00422}{{\tt 1806.00422}}.
%%CITATION = ARXIV:1806.00422;%%.

\bibitem{Tseytlin:1999dj}
A.~A. Tseytlin, ``{Born-Infeld action, supersymmetry and string theory},''
\href{http://www.arXiv.org/abs/hep-th/9908105}{{\tt hep-th/9908105}}.
%%CITATION = HEP-TH/9908105;%%.

\bibitem{Bergshoeff:1986jm}
E.~Bergshoeff, M.~Rakowski, and E.~Sezgin, ``{Higher Derivative SuperYang-Mills
  Theories},'' {\em Phys. Lett.} {\bf B185} (1987)
371--376.
%%CITATION = PHLTA,B185,371;%%.

\bibitem{Metsaev:1987by}
R.~R. Metsaev and M.~Rakhmanov, ``{Fermionic Terms in the Open Superstring
  Effective Action},'' {\em Phys. Lett.} {\bf B193} (1987)
202--206.
%%CITATION = PHLTA,B193,202;%%.

\bibitem{Metsaev:1987qp}
R.~R. Metsaev, M.~Rakhmanov, and A.~A. Tseytlin, ``{The {Born-Infeld} Action as
  the Effective Action in the Open Superstring Theory},'' {\em Phys. Lett.}
  {\bf B193} (1987)
207--212.
%%CITATION = PHLTA,B193,207;%%.

\bibitem{Paban:1998ea}
S.~Paban, S.~Sethi, and M.~Stern, ``{Constraints from extended supersymmetry in
  quantum mechanics},'' {\em Nucl. Phys.} {\bf B534} (1998) 137--154,
\href{http://www.arXiv.org/abs/hep-th/9805018}{{\tt hep-th/9805018}}.
%%CITATION = HEP-TH/9805018;%%.

\bibitem{Paban:1998qy}
S.~Paban, S.~Sethi, and M.~Stern, ``{Supersymmetry and higher derivative terms
  in the effective action of Yang-Mills theories},'' {\em JHEP} {\bf 06} (1998)
  012,
\href{http://www.arXiv.org/abs/hep-th/9806028}{{\tt hep-th/9806028}}.
%%CITATION = HEP-TH/9806028;%%.

\bibitem{Lin:2015ixa}
Y.-H. Lin, S.-H. Shao, Y.~Wang, and X.~Yin, ``{Higher derivative couplings in
  theories with sixteen supersymmetries},'' {\em Phys. Rev.} {\bf D92} (2015),
  no.~12, 125017,
\href{http://www.arXiv.org/abs/1503.02077}{{\tt 1503.02077}}.
%%CITATION = ARXIV:1503.02077;%%.

\bibitem{Chen:2015hpa}
W.-M. Chen, Y.-t. Huang, and C.~Wen, ``{Exact coefficients for higher
  dimensional operators with sixteen supersymmetries},'' {\em JHEP} {\bf 09}
  (2015) 098,
\href{http://www.arXiv.org/abs/1505.07093}{{\tt 1505.07093}}.
%%CITATION = ARXIV:1505.07093;%%.

\bibitem{Garousi:2017fbe}
M.~R. Garousi, ``{Duality constraints on effective actions},'' {\em Phys.
  Rept.} {\bf 702} (2017) 1--30,
\href{http://www.arXiv.org/abs/1702.00191}{{\tt 1702.00191}}.
%%CITATION = ARXIV:1702.00191;%%.

\bibitem{Heydeman:2017yww}
M.~Heydeman, J.~H. Schwarz, and C.~Wen, ``{M5-Brane and D-Brane Scattering
  Amplitudes},'' {\em JHEP} {\bf 12} (2017) 003,
\href{http://www.arXiv.org/abs/1710.02170}{{\tt 1710.02170}}.
%%CITATION = ARXIV:1710.02170;%%.

\bibitem{cecottiSupersymmetricBorninfeldLagrangians1987}
S.~Cecotti and S.~Ferrara, ``Supersymmetric Born-Infeld Lagrangians,'' {\em
  Physics Letters B} {\bf 187} (Mar., 1987) 335--339.

\bibitem{Conti:2018jho}
R.~Conti, L.~Iannella, S.~Negro, and R.~Tateo, ``{Generalised Born-Infeld
  models, Lax operators and the $\textrm{T} \bar{\textrm{T}}$ perturbation},''
\href{http://www.arXiv.org/abs/1806.11515}{{\tt 1806.11515}}.
%%CITATION = ARXIV:1806.11515;%%.

\bibitem{Baggio:2018rpv}
M.~Baggio, A.~Sfondrini, G.~Tartaglino-Mazzucchelli, and H.~Walsh, ``{On
  $T\bar{T}$ deformations and supersymmetry},''
\href{http://www.arXiv.org/abs/1811.00533}{{\tt 1811.00533}}.
%%CITATION = ARXIV:1811.00533;%%.

\bibitem{smirnovSpaceIntegrableQuantum2017}
F.~A. Smirnov and A.~B. Zamolodchikov, ``On Space of Integrable Quantum Field
  Theories,'' {\em Nuclear Physics B} {\bf 915} (Feb., 2017) 363--383,
  \href{http://www.arXiv.org/abs/1608.05499}{{\tt 1608.05499}}.

\bibitem{krausCutoffAdSBarT2018}
P.~Kraus, J.~Liu, and D.~Marolf, ``Cutoff {{AdS}}$ _3$ versus the $T \Tb$
  Deformation,'' {\em Journal of High Energy Physics} {\bf 2018} (July, 2018)
  \href{http://www.arXiv.org/abs/1801.02714}{{\tt 1801.02714}}.

\bibitem{Ferrara:1974pz}
S.~Ferrara and B.~Zumino, ``{Transformation Properties of the Supercurrent},''
  {\em Nucl. Phys.} {\bf B87} (1975)
207.
%%CITATION = NUPHA,B87,207;%%.

\bibitem{dumitrescuSupercurrentsBraneCurrents2011a}
T.~T. Dumitrescu and N.~Seiberg, ``Supercurrents and {{Brane Currents}} in
  {{Diverse Dimensions}},'' {\em Journal of High Energy Physics} {\bf 2011}
  (July, 2011) \href{http://www.arXiv.org/abs/1106.0031}{{\tt 1106.0031}}.

\end{thebibliography}\endgroup

\end{document}